 \documentclass[final,5p,times,twocolumn]{elsarticle}

\usepackage{amssymb}
\usepackage{lipsum}
\usepackage{graphicx}
\usepackage{xurl}
\usepackage{float}

\journal{Physics Letters B}

\begin{document}

\begin{frontmatter}



\title{Using Neural Networks to Accelerate TALYS-2.0 Nuclear Reaction Simulations}


\author[first]{Wilson Lin\corref{cor1}}
\affiliation[first]{organization={Brookhaven National Laboratory},
            addressline={}, 
            city={Upton},
            postcode={11973}, 
            state={NY},
            country={USA}}
\ead{wlin4@bnl.gov}
\author[second]{Catherine E. Apgar}
\affiliation[second]{organization={Los Alamos National Laboratory},
            addressline={}, 
            city={Los Alamos},
            postcode={87545}, 
            state={NM},
            country={USA}}
\ead{ceapgar@lanl.gov}
\author[third,fourth]{Lee A. Bernstein}
\affiliation[third]{organization={Department of Nuclear Engineering, University of California, Berkeley},
            addressline={}, 
            city={Berkeley},
            postcode={94720}, 
            state={CA},
            country={USA}}
\affiliation[fourth]{organization={Lawrence Berkeley National Laboratory},
            addressline={}, 
            city={Berkeley},
            postcode={94720}, 
            state={CA},
            country={USA}}
\ead{labernstein@lbl.gov}
\author[third]{Yun-Hsuan Lee}
\ead{abby6677@berkeley.edu}
\author[fifth]{Alan B. McIntosh}
\affiliation[fifth]{organization={Cyclotron Institute, Texas A\&M University},
            addressline={}, 
            city={College Station},
            postcode={77843}, 
            state={TX},
            country={USA}}
\ead{alanmcintosh@tamu.edu}
\author[first]{Dmitri G. Medvedev}
\ead{dmedvede@bnl.gov}
\author[second]{Ellen M. O'Brien}
\ead{emobrien@lanl.gov}
\author[second]{Christiaan E. Vermeulen}
\ead{etienne@lanl.gov}
\author[third,fourth]{Andrew S. Voyles}
\ead{asvoyles@lbl.gov}
\author[first]{Jonathan T. Morrell}
\ead{jmorrell@bnl.gov}

\cortext[cor1]{Corresponding author.}
\begin{abstract}
Recent efforts to improve the predictability of TALYS-2.0 calculated charged-particle residual product cross sections have focused on adjusting parameters related to the optical model potential and pre-equilibrium process. Although adjusted TALYS-2.0 outputs show marked improvements in agreement with experimental data over the default parameters, the procedure is generally time-consuming due to the need for sequential TALYS-2.0 calculations. Since the models and model parameters must be defined and constrained prior to adjustment, we show in this work that an artificial neural network can serve as a surrogate model to successfully predict TALYS-2.0 outputs within this domain of input parameters. No practical differences were observed in the trained model’s performance between uniform random, Latin hypercube and Sobol sequence sampling for generating the training datasets. Once validated, trained neural network models were used to adjust TALYS-2.0 nuclear model parameters, where a multi-parameter fitting procedure was not only feasible but optimal for this process. The neural network approach is >1000x faster at generating residual product cross sections than using TALYS-2.0 directly, and a high-fidelity surrogate model could be implemented with \textasciitilde{}1500 TALYS-2.0 files to achieve adjusted cross sections comparable to the previous publication.
\end{abstract}




\end{frontmatter}




\section{Introduction}
\label{introduction}
Nuclear transmutations have been an invaluable basis for supplying critical isotopes used in fundamental research \cite{brzezinski_evaluation_1997, shusterman_surprisingly_2019}, medical \cite{engle_recommended_2019,sgouros_radiopharmaceutical_2020} and security \cite{simmonds_design_1967} applications. The availability of high-quality nuclear cross section data is therefore essential for optimizing the production of radioisotopes from various pathways. If experimental data does not exist, nuclear modeling codes serve as a predictive source for isotope production applications \cite{engle_recommended_2019, koning_talys_2023}. These codes often have adjustable parameters that could be tuned to achieve better agreement with experimental data and future generalizability. 

A comprehensive set of charged-particle nuclear cross sections for various target materials of interest to the U.S. Department of Energy Isotope Program were previously measured by the Tri-lab collaboration between Brookhaven National Laboratory, Lawrence Berkeley National Laboratory and Los Alamos National Laboratory \cite{apgar_117msn_2025, fox_investigating_2021, fox_measurement_2021,morrell_measurement_2024}. During the evaluation of these measured cross sections, the authors implemented a parameter adjustment procedure using the TALYS-2.0 \cite{koning_talys_2023} nuclear reaction model code. However, this iterative workflow currently requires sequential TALYS-2.0 calculations, which makes the parameter adjustment process time-consuming and limits parallelization.

Alternatively, simpler models that approximate the complex function of interest, commonly referred to as surrogate models, are often implemented for computationally challenging tasks \cite{kudela_recent_2022}. In building these models, a major advantage is that a small subset of the true model output can potentially represent the problem domain. Since surrogate models evaluate much faster than true models, the total computation time can be significantly reduced. Surrogate modeling has thus been considered in a variety of applications, such as data science, engineering and fundamental physics \cite{alizadeh_managing_2020, cozad_learning_2014, asher_review_2015, mccabe_multiple_2024, samadian_application_2025}. Furthermore, by leveraging recent technological advances in machine learning, surrogate modeling can become more generalizable and effective than simple analytical functions \cite{bosso_application_2024, chahrour_comparing_2022, khan_physics-inspired_2020, khan_ai_2022, trinchero_machine_2021}.

In this work, we demonstrate that the TALYS-2.0 parameter adjustment process can be facilitated by interpolating nuclear model parameters using an artificial neural network (NN) as a surrogate model (see Figure \ref{workflow_fig}). TALYS-2.0 datasets were generated in parallel using either a High-Performance Computing (HPC) cluster at Brookhaven National Laboratory or a standalone Ryzen 7960X CPU. Initially, datasets with three variable parameters were used to characterize the performance of the NN model and enable comparisons with a conventional cubic interpolator. The three parameters chosen here are used in the semi-empirical expression for the squared matrix element, which is relevant to the pre-equilibrium process in TALYS-2.0 \cite{koning_talys_2023}. After evaluating the 3-parameter surrogate model, datasets with six variable parameters were used to test the feasibility of the NN model for scaling up to more practical scenarios. Finally, the NN model was evaluated on datasets varying seventeen parameters matching those previously published for proton-induced reactions on (effectively) La-139 \cite{morrell_measurement_2024}. After validating the performance of the NN model for generating cross sections based on different nuclear model parameters, trained NN models were used to adjust these parameters based on experimental data for \textsuperscript{139}La(p,x) nuclear reactions.

\section{Surrogate modeling using artificial neural networks}

\subsection{TALYS-2.0 cross section dataset generation}
The TALYS-2.0 nuclear reaction model code was used to generate residual product cross section datasets due to its ease of accessibility, suitability for proton induced reactions and ability to use a two-component exciton model for the pre-equilibrium process \cite{morrell_measurement_2024}. For all NN models in this work, the training dataset was uniform random sampled for splitting into training the NN model and for validating the model outputs at each epoch. Test datasets were only used for evaluating the trained model and separated from the training dataset. Proton-induced reactions on \textsuperscript{nat}Cu were initially chosen to balance the available reaction pathways and TALYS-2.0 computation time. Energy sampling similar to TENDL \cite{koning_tendl_2019} was chosen to reduce computation time while retaining sufficient data points to cover the range of interest. Details of the datasets for surrogate models investigated in this work can be found in Table \ref{dataset_tab}.

\subsection{Artificial neural network architecture and training methodology}

Artificial NN models were implemented using PyTorch 2.5.1 \cite{ansel_pytorch_2024} and Python 3.11 \cite{van_rossum_python_2009}. An Intel® Core™ i7-13700 CPU with 64 GB memory was used for establishing the NN model. A single fully connected deep feedforward NN was chosen to establish a working surrogate model for TALYS-2.0. The size of the NN was optimized based on the validation performance, training/computation time and memory constraints (final nodes in hidden layers: 256, 128, 512, 512 and 512). The NN model was trained using the AdamW optimizer \cite{kingma_adam_2017, loshchilov_decoupled_2019} with a Huber loss metric \cite{huber_robust_1964, meyer_alternative_2020}. The learning rate decayed after every 500 epochs, and training was stopped based on the validation performance. Overfitting was mitigated by incorporating weight decay, batch normalization and dropout. The activation function of the input, hidden and output layers were SELU\cite{klambauer_self-normalizing_2017}, Mish \cite{misra_mish_2020} and ReLU \cite{fukushima_visual_1969, nair_rectified_2010}, respectively. 
Residual product cross sections were normalized by the maximum cross section of the respective reaction channel from the default TALYS-2.0 output for training the NN model. 

\subsection{Surrogate model evaluations}
\subsubsection{3-parameter model}
A total of 4096 files were generated via a 16x16x16 grid (16 values for each of the three nuclear parameters) to assess the performance of the 3-parameter surrogate model against a conventional cubic interpolator. The full training dataset (training and validation) for the NN model were provided as sample points for the multidimensional cubic interpolator, and both models were evaluated based on their performance for 30 residual products with the largest integrated cross sections from 0-200 MeV.

\begin{figure}
\includegraphics[width=0.45\textwidth]{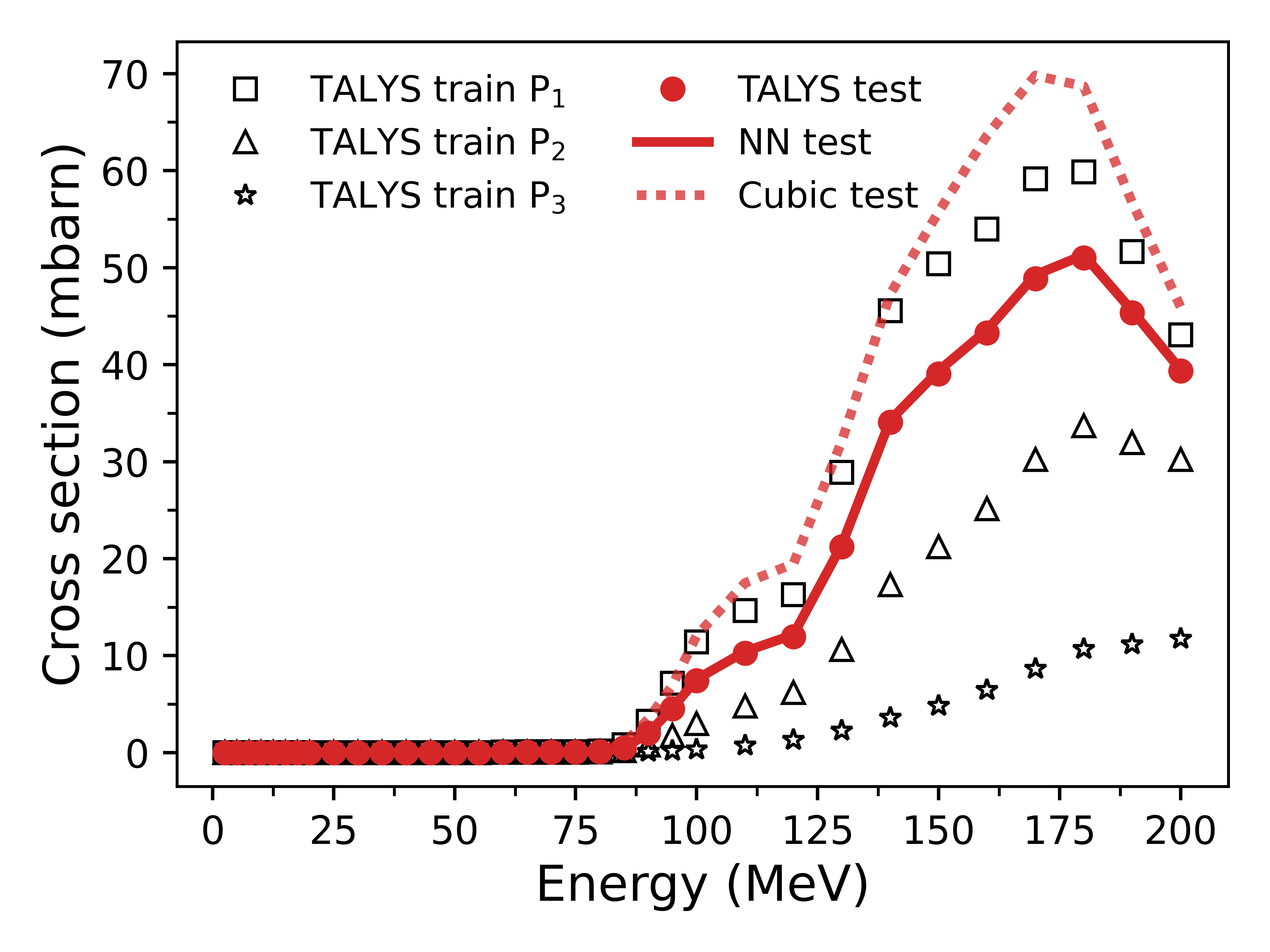}	
	\caption{Comparison between the NN surrogate model (red solid) and standard cubic interpolator (red dotted) after training with 6x6x6 TALYS-2.0 files. The cubic interpolator incurred substantially larger errors than NN for generating the test data (red circle). For reference, the reaction channel shown is \textsuperscript{nat}Cu(p,x)\textsuperscript{51}Cr, and three separate parameter sets used for training both models are plotted as black hollow markers for illustration.} 
	\label{interp_NN_compare_fig}%
\end{figure}

In a direct comparison using 9x9x9 training data (parameter values: 0.20, 0.31, 0.47, 0.72, 1.1, 1.7, 2.6, 4.0 and 5.0), the conventional cubic interpolator was able to achieve a root mean squared error (RMSE) of 0.11 mb and 5.3 mb maximum absolute error (MAE) for the test dataset. Similarly, the NN model achieved 0.24 mb RMSE and 4.7 mb MAE for the test dataset. However, in contrast to the conventional interpolator, the NN model was more robust when trained with a 6x6x6 dataset (parameter values: 0.20, 0.38, 0.72, 1.4, 2.6, 5.0), achieving 0.22 mb RMSE (0.89 mb interpolator) and 5.6 mb MAE (50 mb interpolator). This performance difference is illustrated in Figure \ref{interp_NN_compare_fig}, where both models trained with 6x6x6 TALYS-2.0 files were used to generate cross sections from the test dataset. These results suggest that the NN model can achieve good agreement with TALYS-2.0 generated cross sections for small training datasets.

\subsubsection{6-parameter model}
With 6 nuclear model parameters and the increased test dataset size (\textasciitilde{}65k files), the conventional cubic interpolator was unable to compute \textasciitilde{}66 million cross sections within a reasonable amount of time (>1 day). Alternatively, the NN model could predict the cross sections in <1 min and achieved 0.71 mb RMSE and 49 mb MAE with just 300 TALYS-2.0 files for the training dataset. The hyperparameters obtained from evaluating the 3-parameter NN model required only minor adjustments for optimization. However, likely due to the increased dimensionality and larger test dataset size, approximately twice as many training files were needed to achieve comparable RMSE to the 3-parameter NN model.

\subsubsection{17-parameter model}
Continuing from promising results with the 6-parameter NN model, the surrogate modeling was expanded to interpolate 17 nuclear model parameters. These parameters and associated ranges were previously selected by Morrell et al. \cite{morrell_measurement_2024} to improve agreement between TALYS-2.0 outputs and experimental measurements for \textsuperscript{139}La(p,x). The reaction cross sections were calculated via TALYS-2.0 using the same model selections (ldmodel 5, strength 10, deuteronomp 2, alphaomp 4, preeqspin 3). Short-lived parents of residual products from reactions used in the parameter adjustment procedure were also included in the NN model. As such, this scenario serves to illustrate a realistic application of the TALYS-2.0 surrogate model approach for nuclear model parameter adjustments. Three different sampling methods (uniform random, Latin hypercube \cite{loh_latin_1996,mckay_comparison_1979,stein_large_1987} and shuffled Sobol sequence \cite{sobol_distribution_1967, owen_scrambling_1998, joe_constructing_2008}) were selected for the training dataset to assess potential differences in the performance of the trained NN model.

\begin{figure}
	\includegraphics[width=.45\textwidth]{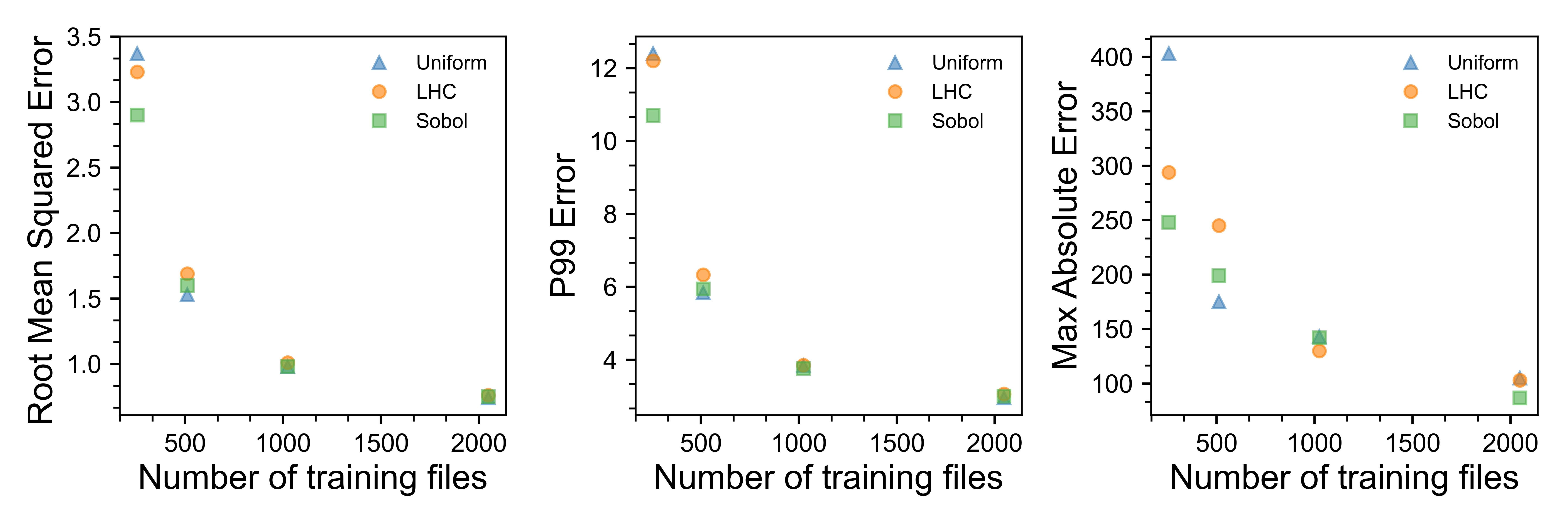}	
	\caption{Different metrics used to evaluate the performance of the surrogate model trained using files sampled from uniform (blue triangle), Latin hypercube (LHC, orange circle) or Sobol sequence (green square) sampling. All three sampling methods are generally comparable and show a general trend of improved performance with increasing training dataset size.} 
	\label{model_FOM_fig}%
\end{figure}

Based on RMSE, 99th percentile absolute error (P99) and MAE, NN models trained with different sampling methods were roughly comparable with the same number of training files (Figure \ref{model_FOM_fig}, Figure \ref{linreg_la_fig} and Table \ref{metric_tab}). Larger training datasets generally achieved better performance and no observable differences in individual reaction channels were observed for the different sampling methods (Sobol: Figure \ref{sobol_la_fig}, Latin hypercube: Figure \ref{lhc_la_fig}, and uniform: Figure \ref{uni_la_fig}).

To verify the robustness of the NN approach for other target material, \textsuperscript{nat}Cu(p,x) nuclear reactions were assessed using the same set of nuclear model parameters, physics model selection, NN architecture/hyperparameters and training dataset size. A set of 26 reaction products were chosen to span cross sections from 0.1-500 mb. The NN model’s performance was also evaluated using a test dataset of 25k uniform random sampled files within the specified parameter bounds. Without any modifications to the hyperparameters, the NN model for \textsuperscript{nat}Cu(p,x) achieved similar results to \textsuperscript{139}La(p,x) with similarly sized training datasets (Figure \ref{linreg_cu_fig}). 

\section{Parameter adjustment using the trained neural network}
After validating the NN model’s ability to predict cross sections by interpolating outputs of different nuclear model parameters, the best performing NN model in this work (Sobol with 2048 files) was chosen to replicate the parameter adjustment procedure from \cite{morrell_measurement_2024}. Each parameter was optimized sequentially starting from “rvadjust n” to “Cstrip a” and repeated until convergence (i.e., a “1-D” adjustment procedure). Since the convergence criteria was not explicitly provided, this was set as 0.001 relative difference since the last full set of parameter iteration. The chi-squared figure of merit (FOM) was weighted by the average fractional contributions of the integral cross sections from 0-200 MeV and the maximum cross section of each reaction product based on default TALYS-2.0. Also, maintaining the same procedure as before, the weight for Ce-134 was increased by a factor of 2 due to its potential as a diagnostic congener for Ac-225 radiopharmaceuticals. The NN model was used to generate 100 datapoints for each parameter with (parameter scaled) linear spacing. The resulting best-fit parameters were then used in TALYS-2.0 calculations to verify the NN model. Experimental cross section data used for parameter adjustment were taken from \cite{becker_cross_2020, morrell_measurement_2020, morrell_measurement_2024, tarkanyi_activation_2017}, where data with no reported uncertainties or uncertainties <5\% were set to 5\%.

The 1-D parameter adjustment procedure using the NN model achieved comparable results to \cite{morrell_measurement_2024} (Figure \ref{adj_1d_fig}), and the NN generated cross sections mostly agree with TALYS-2.0 (Figure \ref{adj_rdiff_fig}). However, despite the comparable residual product cross sections, several parameters were noticeably different from the previously determined set of parameters by \cite{morrell_measurement_2024} (Table \ref{adj_param_vals_tab}). These discrepancies could be explained by the lack of sensitivity for several parameters (Figure \ref{FOM_1d_fig}) and slight differences in the convergence criteria and experimental uncertainty data. Several of the 1-D adjusted parameters are also at the ends of their respective ranges, which could contribute to greater deviations from the true TALYS-2.0 outputs since less training data are expected to exist at these extremes.

\begin{table*}
\centering
\begin{tabular}{l c c c c c} 
 \hline
 Parameter & Range & Morrell et al. \cite{morrell_measurement_2024} & This work 1-D & This work N-D & This work N-D \\ 
        & & & Sobol & Sobol & Uniform \\
        & & & 2048 & 2048 & 1536 \\
 \hline
rvadjust   n & 0.95-1.05 & 0.977  & 0.991 & 0.984 & 0.995 \\
avadjust n   & 0.75-1.25 & 1.076  & 1.018 & 1.039 & 0.918 \\
rvadjust p   & 0.9-1.1   & 0.9    & 1.08  & 1.095 & 1.077 \\
avadjust p   & 0.7-1.3   & 1.0118 & 0.7   & 0.733 & 0.729 \\
w1adjust n   & 0.25-4    & 4      & 4     & 3.979 & 3.911 \\
w2adjust n   & 0.25-4    & 1.51   & 1.955 & 1.131 & 1.452 \\
w1adjust p   & 0.25-4    & 4      & 4     & 3.981 & 3.536 \\
w2adjust p   & 0.25-4    & 1.83   & 2.523 & 2.727 & 2.229 \\
M2constant   & 0.2-5     & 1.055  & 0.2   & 0.764 & 0.335 \\
M2shift      & 0.2-5     & 1.196  & 1.412 & 1.912 & 1.04  \\
M2limit      & 0.2-5     & 1.926  & 1.412 & 0.675 & 1.569 \\
Rnunu        & 0.1-10    & 1.66   & 10    & 3.592 & 6.866 \\
Rnupi        & 0.1-10    & 0.16   & 4.5   & 3.16  & 1.608 \\
Rpinu        & 0.1-10    & 2.21   & 10    & 9.864 & 5.721 \\
Rpipi        & 0.1-10    & 0.1    & 0.1   & 0.16  & 0.112 \\
Rgamma       & 0.1-10    & 0.1    & 2     & 9.513 & 2.671 \\
Cstrip a     & 0.1-10    & 0.1    & 0.1   & 4.342 & 3.428 \\
\hline
\end{tabular}
\caption{Adjusted nuclear model parameters in TALYS-2.0 using the NN models. The parameter set determined from by Morrell et al. \cite{morrell_measurement_2024} is also presented for comparison, where clear discrepancies exist for several parameters.}
\label{adj_param_vals_tab}
\end{table*}

\begin{figure}
	\includegraphics[width=.45\textwidth]{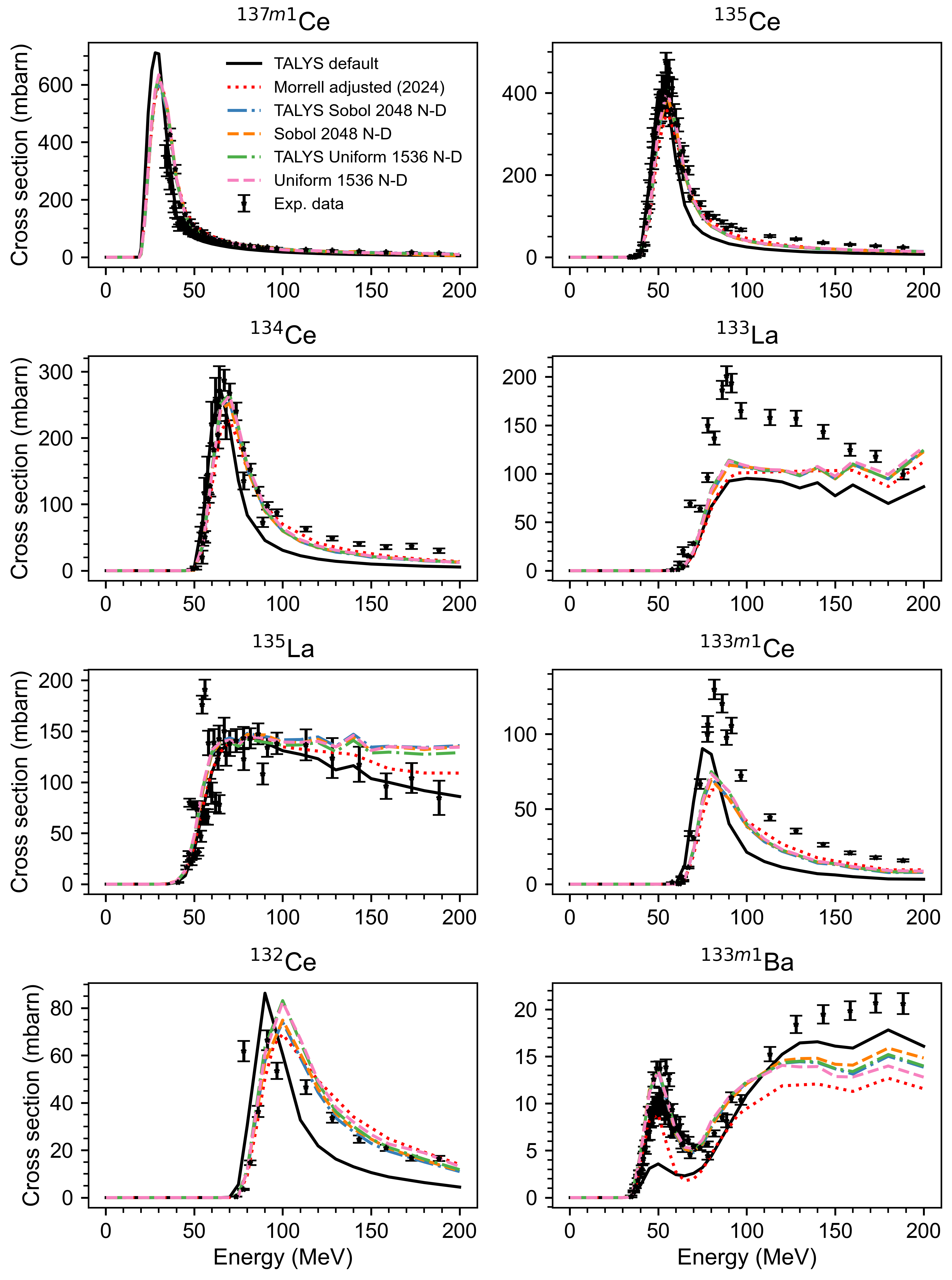}	
	\caption{Resulting excitation functions after the N-D parameter adjustment procedure using the NN model trained with either 2048 files from Sobol sequence (orange dashed) or 1536 files from uniform random sampling (pink dashed). The respective TALYS-2.0 cross sections (dash dotted), experimental data (black asterisk), default TALYS-2.0 (black solid) and previously adjusted output by \cite{morrell_measurement_2024} (red dotted) are shown for comparison.} 
	\label{adj_ND_fig}%
\end{figure}

Alternatively, since the NN surrogate model effectively approximates TALYS-2.0, a multiparameter (“N-D”) adjustment procedure was also implemented to assess potential differences in the two procedures. For the N-D procedure, a dual annealing algorithm \cite{xiang_generalized_1997, xiang_efficiency_2000, xiang_generalized_2013} was used to perform a constrained optimization for the parameters (iterated 10 times). In comparison to the 1-D procedure, the N-D procedure took about 8 min to complete (3 min for 1-D) but achieved better agreement with TALYS-2.0 (Figure \ref{adj_rdiff_fig}) and slightly improved FOM (Figure \ref{adj_ND_fig}). 

\section{Discussion}
The primary aim of this work was to demonstrate that the nuclear model parameter adjustment procedure could be parallelized to benefit from manycore processors/HPC systems. Artificial NN were identified as potential surrogate models for TALYS-2.0 due to their good generalization for complex and high-dimensional tasks \cite{ghorbani_linearized_2021, hastie_surprises_2022, liang_just_2020}. Training datasets for the NN models can be generated in parallel and even prior to experimental evaluations, but physics models and their associated parameters must be clearly identified. In this work, the optimal number of training files for adjusting 17 constrained parameters using a NN model was between 1024 to 2048 TALYS-2.0 output files. Using only a standard Intel i7-13700 CPU for training the NN model (no GPU boost) and an AMD Ryzen 7960X CPU for generating TALYS-2.0 files, a TALYS-2.0 surrogate model can be implemented in <24 h for \textsuperscript{139}La(p,x). Since both the dataset generation and NN training are parallelizable, access to modern HPC clusters and GPU(s), respectively, can further reduce the total processing time to a few hours, which would be 10-100x faster than the sequential parameter adjustment approach. Hyperparameters for the 17-parameter NN model were also robust for \textsuperscript{nat}Cu(p,x), suggesting that this approach could be generalized for other target nuclei. Additionally, the corresponding lightweight surrogate models enable FOM sensitivity analysis and conventional global optimization techniques that would otherwise be impractical.

A (shuffled) Sobol sequence sampling was hypothesized to more effectively train the NN over uniform random and Latin hypercube sampling due to its favorable coverage of multi-dimensional space (Figure \ref{sample_scatter_fig}). Based on the metrics used in this work, Sobol sequence sampling was practically indistinguishable from the simple uniform random sampling method and requires sampling files according to a base power of 2. In a direct comparison, an NN model trained using 1536 uniform random sampled files could achieve similar results to one trained with 2048 files from Sobol sequence sampling (Figure \ref{adj_ND_fig} and Table \ref{adj_FOM_tab}). Similar results were also observed for 1536 Latin hypercube-sampled training files.

The parameter adjustment process resulted in an improved fit to the experimental data for all methods investigated in this work compared to default TALYS-2.0. Larger discrepancies between the NN model and TALYS-2.0 were observed for adjusting parameters with the 1-D approach compared to N-D. This behavior is likely due to the lack of training data near the extreme bounds of the parameter values, where the 1-D procedure is more likely to compound these extremes since the process does not vary parameters in multidimensional space. Even with a relatively strict convergence criteria for the N-D adjustment, the process converged in 8 min with over 100k parameter set evaluations which is >1000x faster than calculating cross sections using TALYS-2.0. Additionally, another major benefit of the surrogate model approach is that the parameter adjustment process itself, such as changing the experimental dataset/FOM and down selecting parameters/reaction channels, can be altered without recalculating TALYS-2.0 datasets. The resulting FOM from the N-D adjustment was also slightly better than both 1-D and Morrell et al. \cite{morrell_measurement_2024}, though improvements to FOM were not the focus of this work.

Future work aims to further reduce the number of files needed to implement a high-fidelity TALYS-2.0 surrogate model, either by optimizing the NN architecture or through different machine learning techniques. The NN approach could also be expanded to incorporate several target nuclei for the same physical models and parameters, where a single multi-target nuclei model would likely require fewer training files than multiple standalone models. Although not explored in this work, an independent module to upscale the incident particle energies per reaction and excitation energy bins would also greatly reduce the computation time for all TALYS-2.0 related tasks.

\section{Summary and conclusions}

A TALYS-2.0 surrogate model was implemented by using artificial neural networks to generate residual product cross sections for at least 17 variable nuclear model parameters. High fidelity TALYS-2.0 surrogate models were established by training with \textasciitilde{}1500 TALYS-2.0 output files and achieved >1000x speedup in computation time for generating nuclear cross sections. Hyperparameters used for training \textsuperscript{139}La(p,x) models were also validated for \textsuperscript{nat}Cu(p,x), suggesting generalizability for other target nuclei. A multi-parameter dual (simulated) annealing optimization resulted in lower deviations between TALYS-2.0 and the surrogate model with improved fitting to experimental data. Future work aims to further reduce the number of training files needed to establish a TALYS-2.0 surrogate model and to expand the approach for a wider range of target nuclei.

\section*{Acknowledgements}

Brookhaven National Laboratory (BNL) is operated by Brookhaven Science Associates, LLC, (BSA) under a contract with the US Department of Energy (DOE, Contract No. DE-SC0012704). This research is supported by the U.S. Department of Energy Isotope Program, managed by the Office of Science for Isotope R\&D and Production. We are grateful to the Computing and Data Sciences directorate at BNL for providing access to HPC1 which helped generate data for initial model evaluations. The authors acknowledge all members of the BNL Isotope Research and Production department for their inputs.
\appendix
\section{Supplementary materials}
\setcounter{figure}{0}


\begin{figure}[H]
\includegraphics[height=.45\textwidth, angle=90, keepaspectratio]{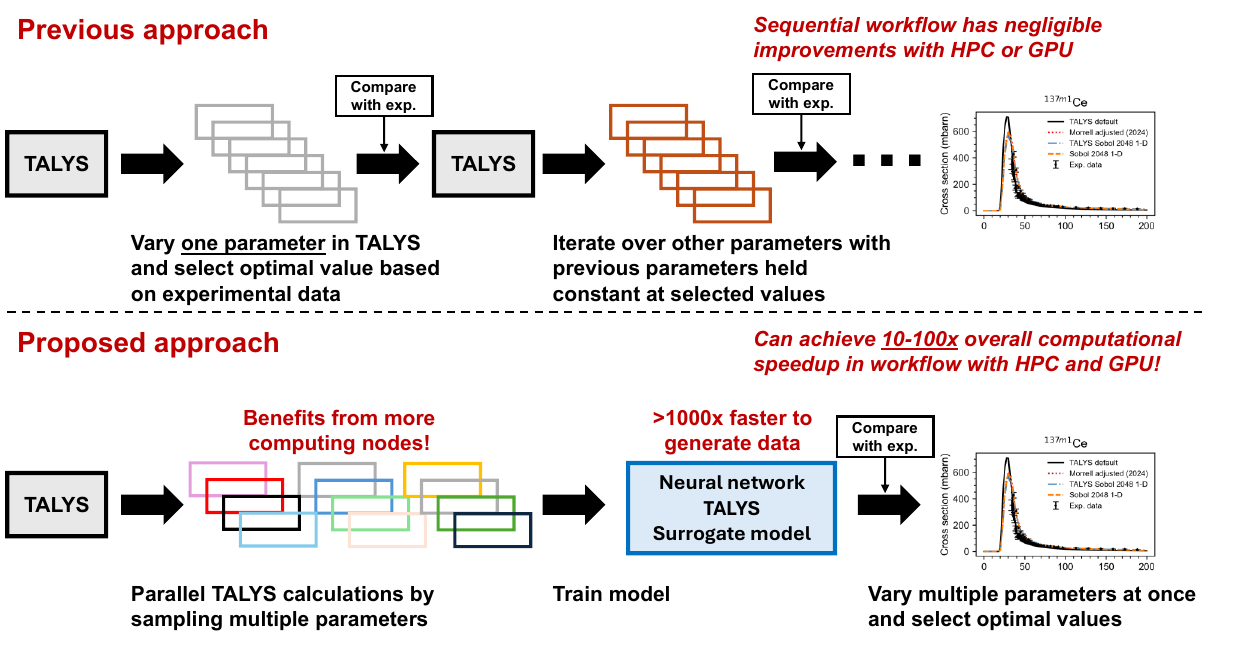}	
	\caption{Comparison of the previous approach by Morrell et al. \cite{morrell_measurement_2024} and the proposed approach in this work for adjusting TALYS-2.0 nuclear model parameters.} 
	\label{workflow_fig}%
\end{figure}

\begin{figure}
\includegraphics[width=.425\textwidth]{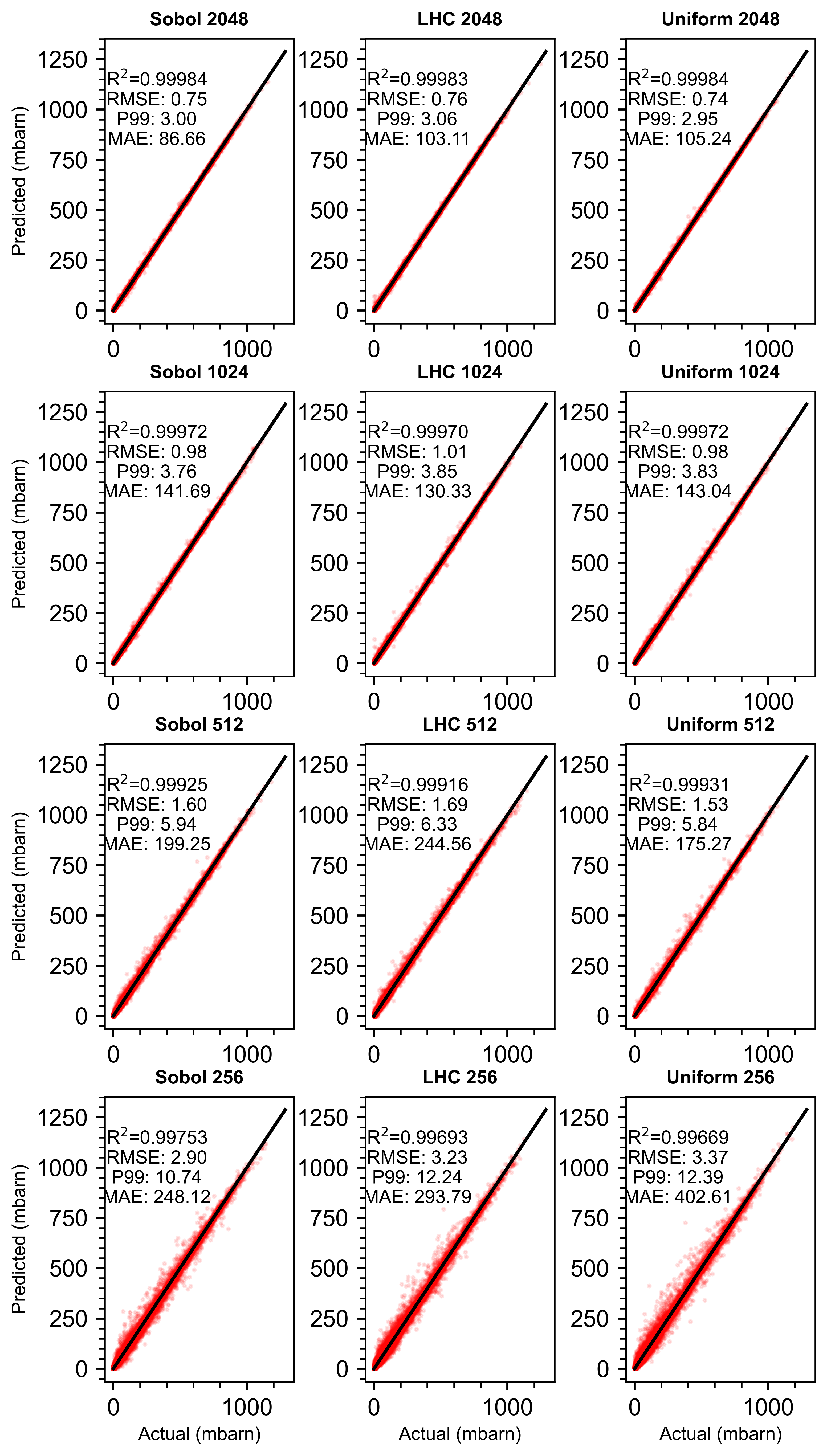}	
	\caption{Actual (TALYS-2.0) vs predicted (NN model) plots for each of the sampling methods and respective number of training files used to train the NN model for \textsuperscript{139}La(p,x). The TALYS-2.0 cross section data were taken from 25k uniform random sampled files not used for training. Due to the large amount of data, only 750k random cross sections were selected for these plots.} 
	\label{linreg_la_fig}%
\end{figure}

\begin{figure}
\includegraphics[width=.45\textwidth]{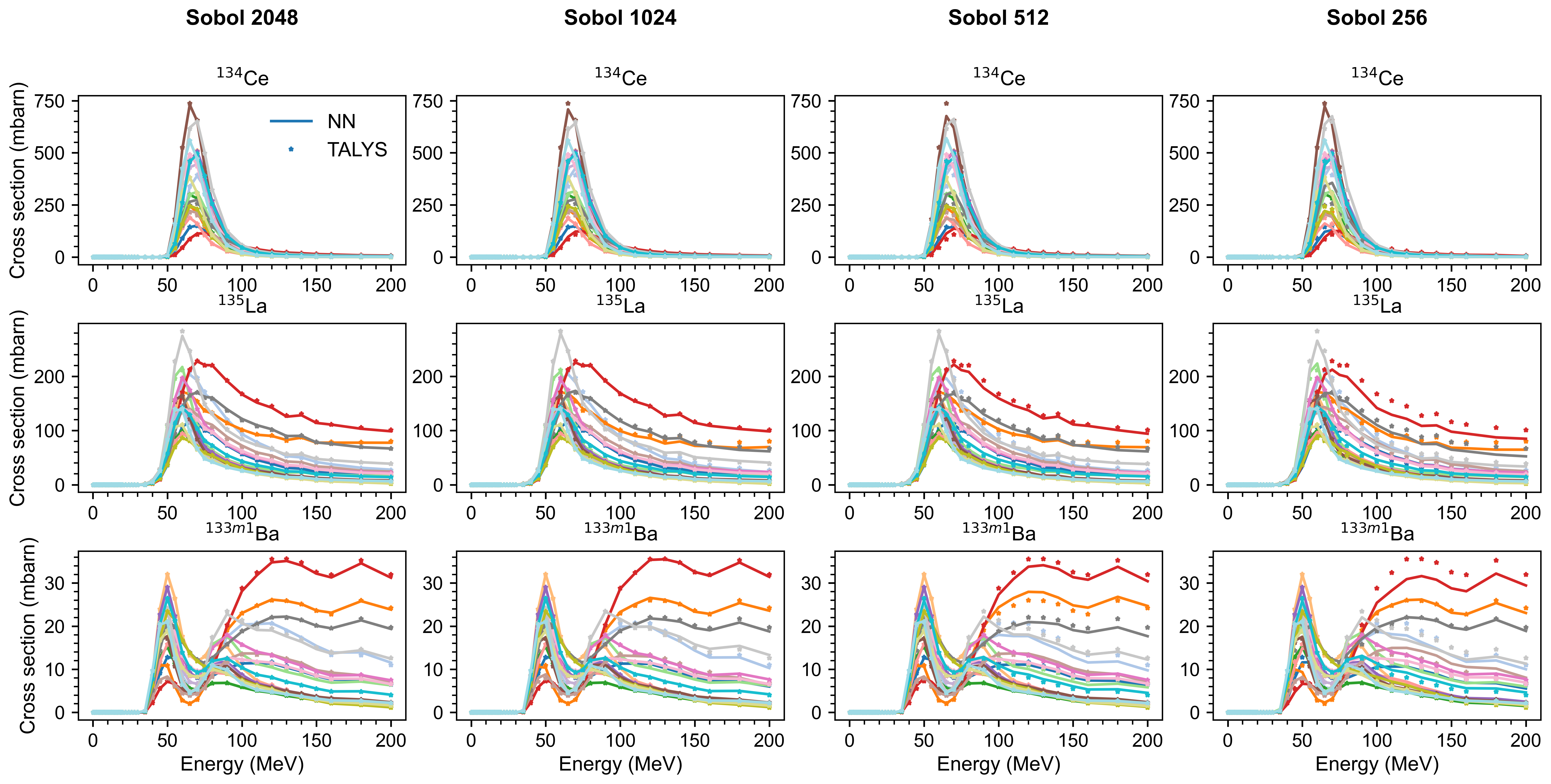}	
	\caption{Sample cross sections predicted by the NN model for \textsuperscript{139}La(p,x), where each column represents a different training dataset sampled by the Sobol sequence. Each color indicates a separate set of nuclear model parameters which were randomly chosen from the test dataset. The NN model generated cross sections (NN, solid line) agree well with those from TALYS-2.0 (asterisks) using 2048 training files, where smaller training datasets begin to incur larger deviations. The NN model generated cross sections have the same energies as TALYS-2.0 and were linearly interpolated for better visibility.} 
	\label{sobol_la_fig}%
\end{figure}

\begin{figure}
\includegraphics[width=.45\textwidth]{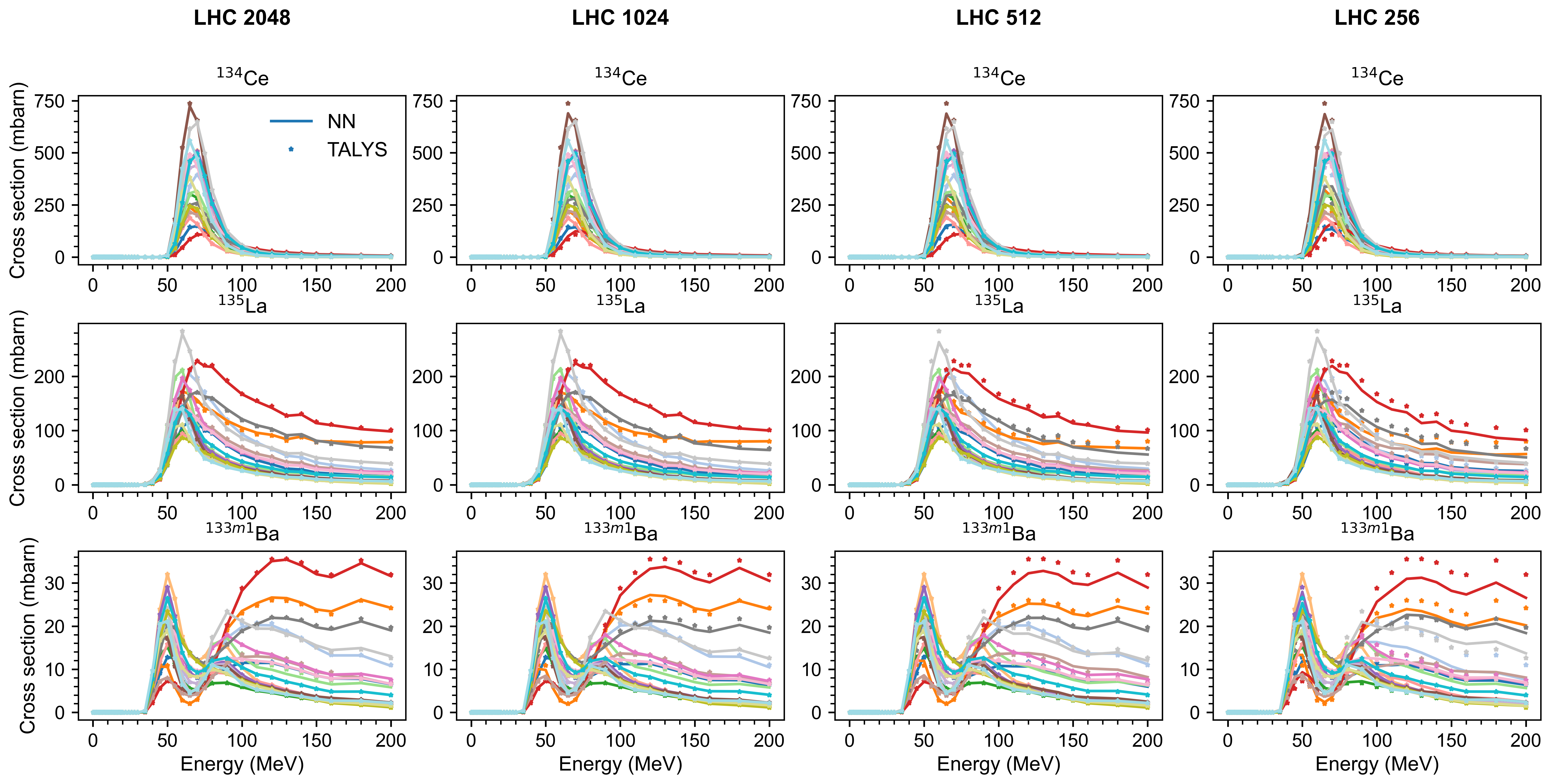}	
	\caption{Sample \textsuperscript{139}La(p,x) cross sections predicted by the NN model for Latin hypercube sampled training datasets.} 
	\label{lhc_la_fig}%
\end{figure}

\begin{figure}
\includegraphics[width=.45\textwidth]{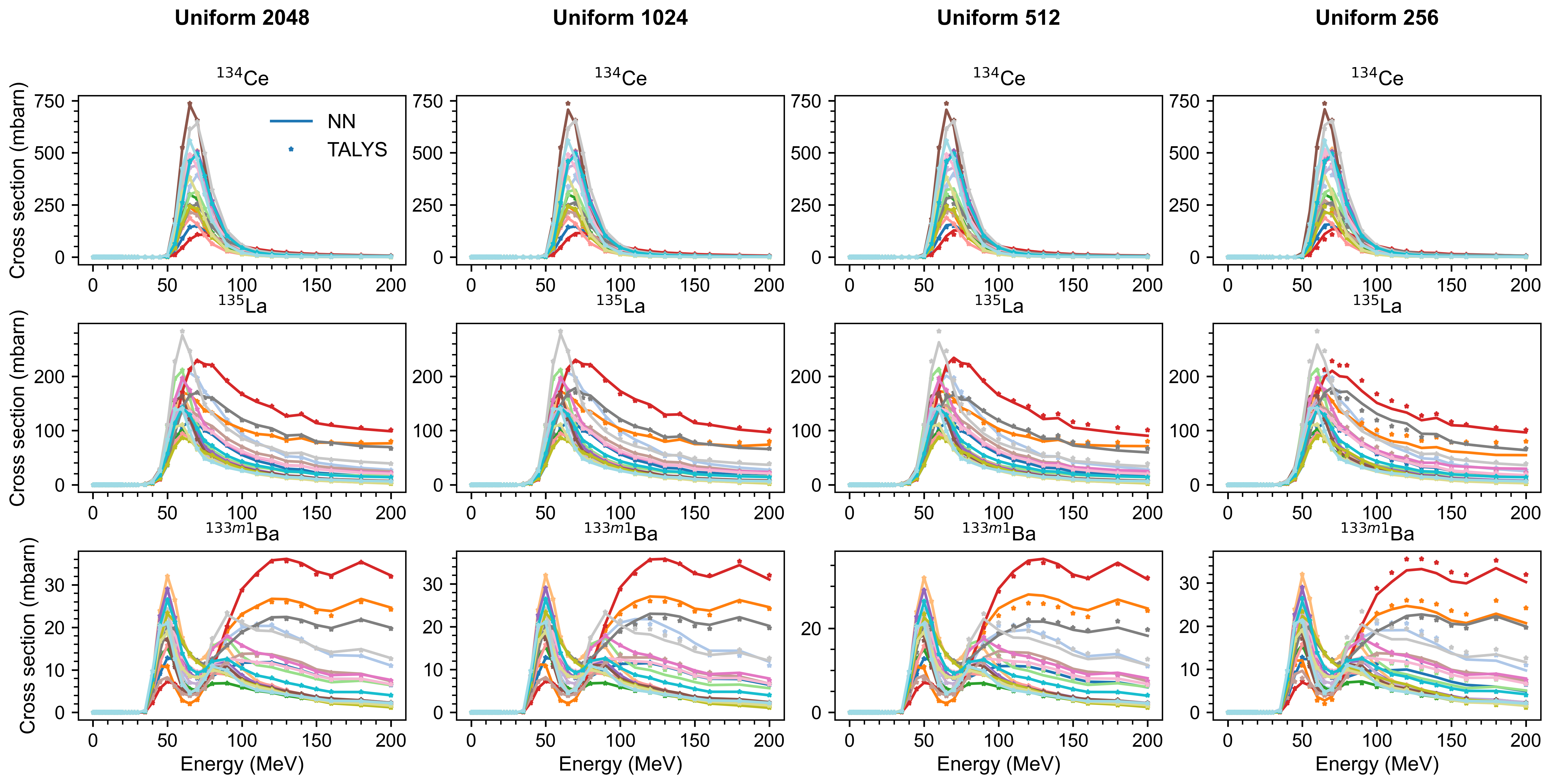}	
	\caption{Sample \textsuperscript{139}La(p,x) cross sections predicted by the NN model for uniform random sampled training datasets.} 
	\label{uni_la_fig}%
\end{figure}

\begin{figure}
\includegraphics[width=.45\textwidth]{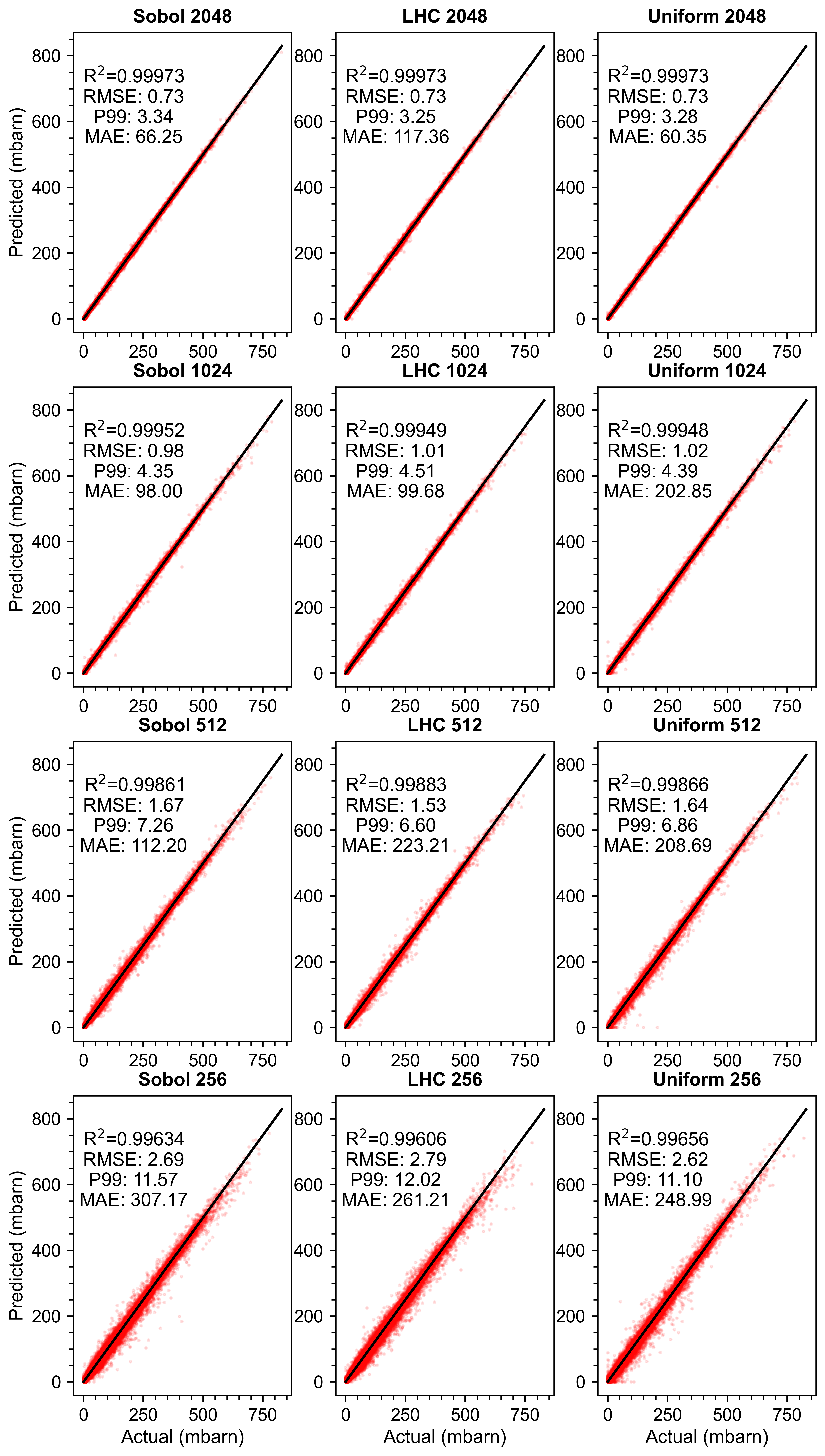}	
	\caption{Actual (TALYS-2.0) vs predicted (NN model) plots for each of the sampling methods and respective number of training files used to train the NN model for \textsuperscript{nat}Cu(p,x).} 
	\label{linreg_cu_fig}%
\end{figure}

\begin{figure}
\includegraphics[width=.45\textwidth]{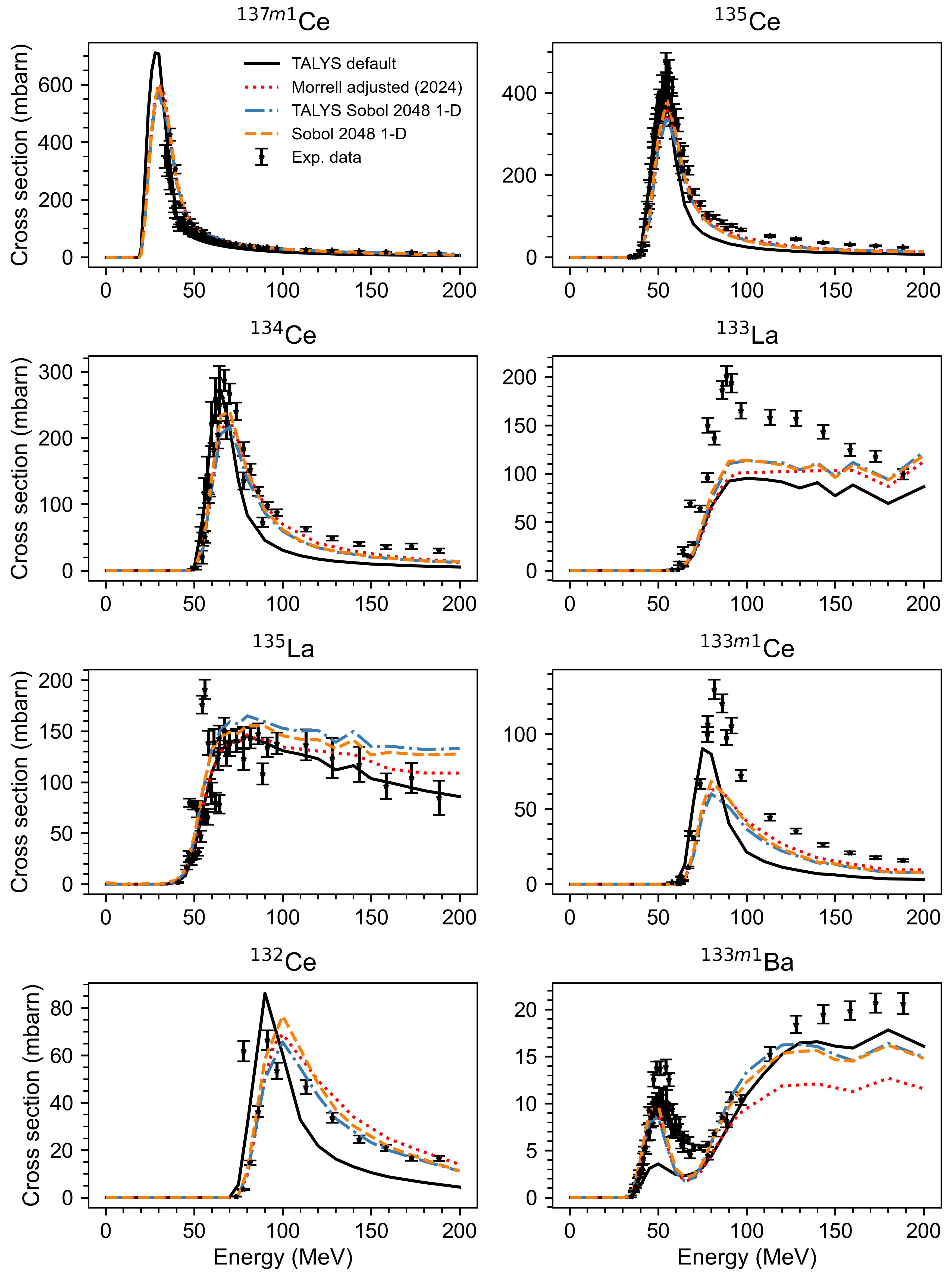}	
	\caption{Excitation functions of several \textsuperscript{139}La(p,x) residual products after performing a 1-D parameter adjustment procedure using the NN model (orange dashed) trained with 2048 Sobol sequenced files. The adjusted parameters were also used for TALYS-2.0 calculations (blue dash dotted) to verify the NN model’s outputs. The default TALYS-2.0 output (black solid) and previously published results from Morrell et al. \cite{morrell_measurement_2024} (red dotted) are also provided for reference.} 
	\label{adj_1d_fig}%
\end{figure}

\begin{figure}
\includegraphics[width=.45\textwidth]{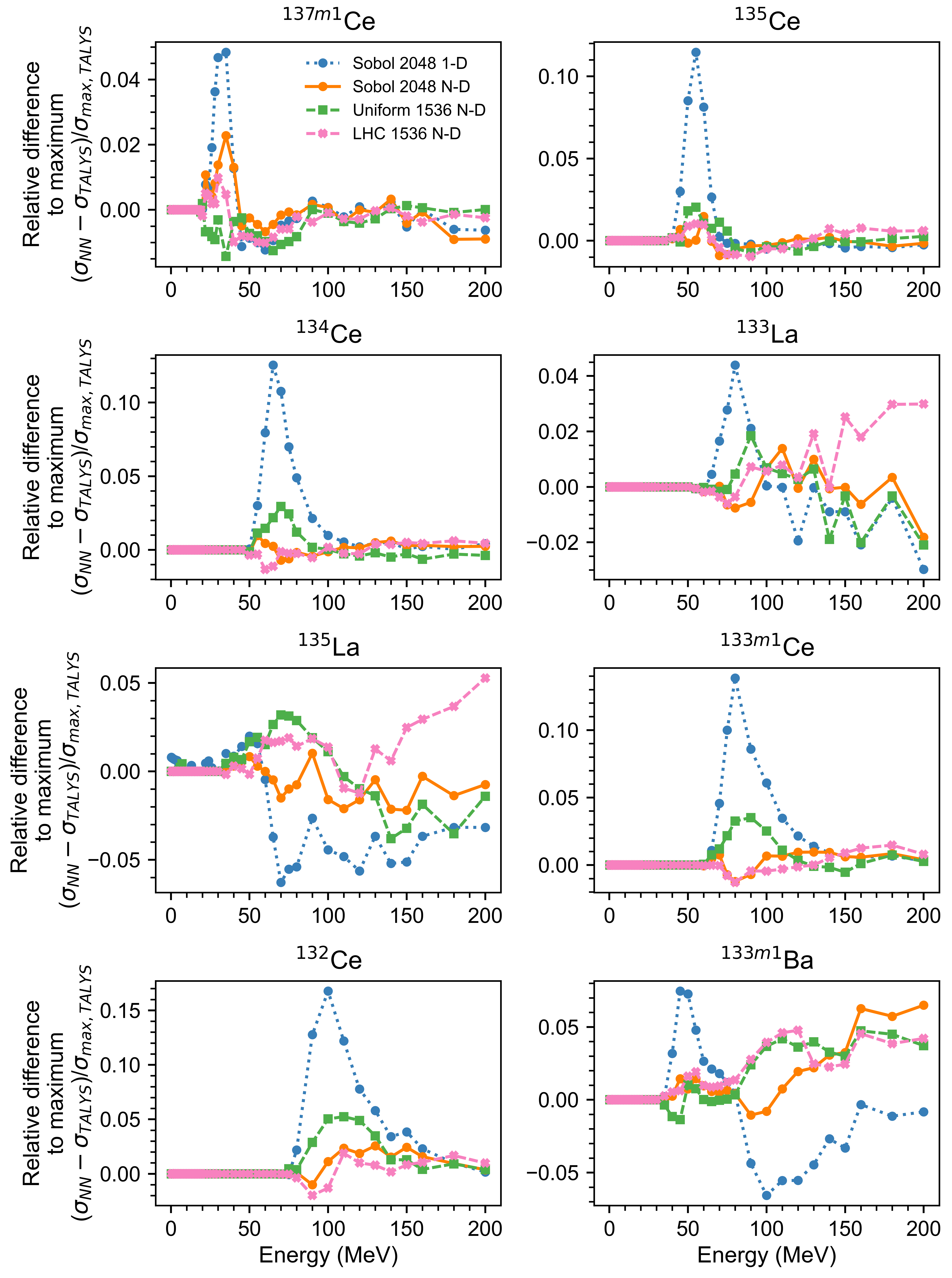}	
	\caption{Deviations of NN generated cross sections from TALYS-2.0 scaled by the maximum TALYS-2.0 cross section. The 1-D parameter adjustment (blue dotted circle) incurred larger deviations than N-D using 2048 Sobol sequence (orange solid circle), 1536 uniform random (green dashed square) or 1536 Latin hypercube (pink dashed X) sampled training files.} 
	\label{adj_rdiff_fig}%
\end{figure}

\begin{figure}
\includegraphics[width=.45\textwidth]{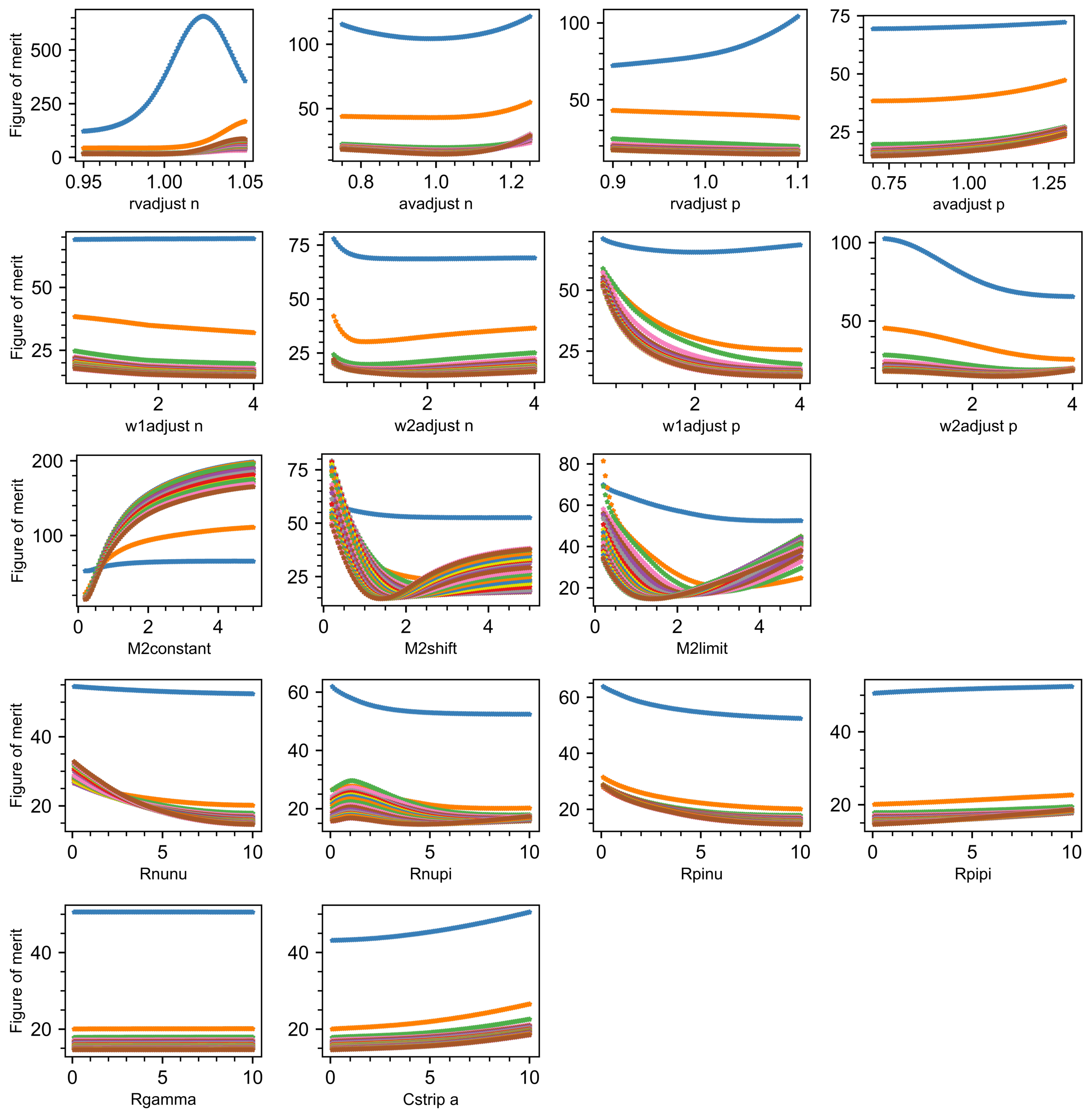}	
	\caption{The impact of each parameter on FOM used in this work for the 1-D parameter adjustment procedure. The NN model was trained with 2048 files from Sobol sequence sampling. Convergence was reached after 23 iterations of the entire parameter set. Each separate color represents one full iteration over the parameters.} 
	\label{FOM_1d_fig}%
\end{figure}

\begin{figure}
\includegraphics[width=.45\textwidth]{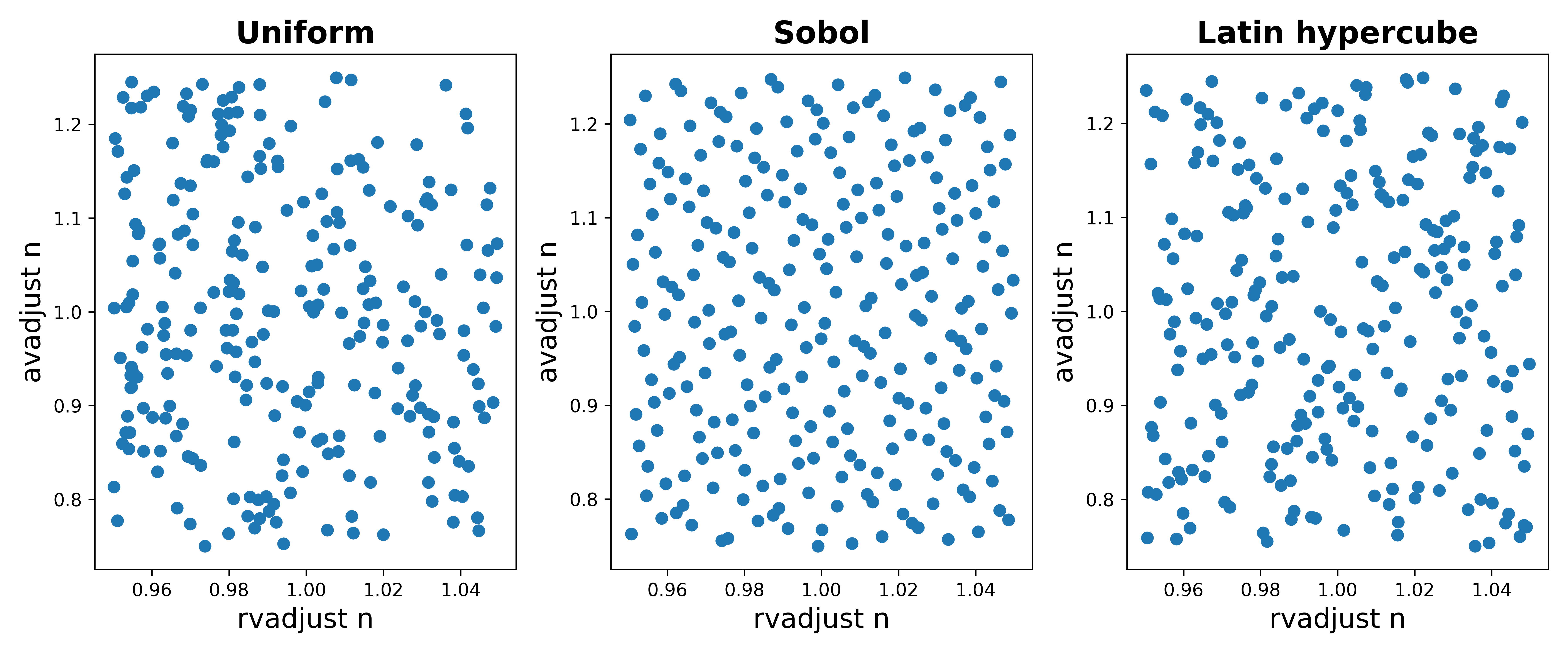}	
	\caption{A comparison of the different sampling methods for the files used in training the NN model (256 files). For simplicity, only two parameters (rvadjust n and avadjust n) were chosen to demonstrate the differences between each sampling method.} 
	\label{sample_scatter_fig}%
\end{figure}

\setcounter{table}{0}


\begin{table}
\begin{tabular}{c c c c} 
 \hline
 Parameters & Training & Test & Number of \\
  & data split & sample method & test files \\
\hline
3  & 50\% & Equidistant log & 4096   \\
6  & 90\% & uniform         & 65000     \\
17 & 90\% & uniform         & 25000     \\
\hline
\end{tabular}
\caption{Additional information related to the datasets used for the surrogate models investigated in this work. The 6-parameter surrogate model incorporated all parameters from the 3-parameter surrogate model, and the 17-model parameter similarly includes all parameters from the 6-parameter surrogate model.}
\label{dataset_tab}
\end{table}

\begin{table}
\begin{tabular}{l c c c c} 
 \hline
 Training files & Metric & \multicolumn{3}{c}{Sampling method} \\
  & & Sobol & LHC & Uniform \\
\hline
 256  & RMSE & 2.9  & 3.23 & 3.37 \\
     & P99  & 10.7 & 12.2 & 12.4 \\
     & MAE  & 248  & 294  & 403  \\
512  & RMSE & 1.6  & 1.69 & 1.53 \\
     & P99  & 5.94 & 6.33 & 5.84 \\
     & MAE  & 199  & 245  & 175  \\
1024 & RMSE & 0.98 & 1.01 & 0.98 \\
     & P99  & 3.76 & 3.85 & 3.83 \\
     & MAE  & 142  & 130  & 143  \\
2048 & RMSE & 0.75 & 0.76 & 0.74 \\
     & P99  & 3    & 3.06 & 2.95 \\
     & MAE  & 86.7 & 103  & 105 
\\ 
 \hline
\end{tabular}
\caption{Metrics used to evaluate the NN model’s performance with different training datasets for \textsuperscript{139}La(p,x). The same test dataset consisting of 25k uniform random sampled files was used for all evaluations.}
\label{metric_tab}
\end{table}

\begin{table}
\begin{tabular}{l c c c c} 
 \hline
Training & RMSE & P99 & MAE & FOM  \\
files & (mb) & (mb) & (mb) & \\
\hline
Uniform 1536   & 0.77      & 3.03     & 123.7    & 14.4 \\
LHC 1536 & 0.79 & 3.10 & 121.6 & 14.0 \\
Sobol 2048     & 0.75      & 3        & 86.7     & 14.6 \\ 
 \hline
\end{tabular}
\caption{Comparison between NN models trained with either 1536 uniform random/LHC or 2048 Sobol sequence sampled files. FOM for the surrogate models were determined using the ND parameter adjustment procedure. For reference, the FOM for default TALYS-2.0 and Morrell et al. \cite{morrell_measurement_2024} was 73.7 and 16.1, respectively. }
\label{adj_FOM_tab}
\end{table}


\bibliographystyle{ieeetr} 
\bibliography{example}

@article{koning_talys_2023,
	title = {{TALYS}: modeling of nuclear reactions},
	volume = {59},
	issn = {1434-601X},
	url = {https://doi.org/10.1140/epja/s10050-023-01034-3},
	doi = {10.1140/epja/s10050-023-01034-3},
	shorttitle = {{TALYS}},
	pages = {131},
	number = {6},
	journaltitle = {The European Physical Journal A},
	shortjournal = {Eur. Phys. J. A},
	author = {Koning, Arjan and Hilaire, Stephane and Goriely, Stephane},
	urldate = {2024-10-31},
	date = {2023-06-14},
	langid = {english},
}

@misc{morrell_measurement_2024,
	title = {Measurement of Proton-Induced Reactions on Lanthanum from 55--200 {MeV} by Stacked-Foil Activation},
	url = {http://arxiv.org/abs/2402.17893},
	number = {{arXiv}:2402.17893},
	publisher = {{arXiv}},
	author = {Morrell, Jonathan T. and O'Brien, Ellen M. and Skulski, Michael and Voyles, Andrew S. and Medvedev, Dmitri G. and Mocko, Veronika and Bernstein, Lee A. and Vermeulen, C. Etienne},
	urldate = {2024-10-31},
	date = {2024-02-27},
	langid = {english},
	eprinttype = {arxiv},
	eprint = {2402.17893 [nucl-ex]},
}

@article{fox_measurement_2021,
	title = {Measurement and modeling of proton-induced reactions on arsenic from 35 to 200 {MeV}},
	volume = {104},
	url = {https://link.aps.org/doi/10.1103/PhysRevC.104.064615},
	doi = {10.1103/PhysRevC.104.064615},
	pages = {064615},
	number = {6},
	journaltitle = {Physical Review C},
	shortjournal = {Phys. Rev. C},
	publisher = {American Physical Society},
	author = {Fox, Morgan B. and Voyles, Andrew S. and Morrell, Jonathan T. and Bernstein, Lee A. and Batchelder, Jon C. and Birnbaum, Eva R. and Cutler, Cathy S. and Koning, Arjan J. and Lewis, Amanda M. and Medvedev, Dmitri G. and Nortier, Francois M. and O'Brien, Ellen M. and Vermeulen, Christiaan},
	urldate = {2024-11-13},
	date = {2021-12-27},
}

@article{fox_investigating_2021,
	title = {Investigating high-energy proton-induced reactions on spherical nuclei: Implications for the preequilibrium exciton model},
	volume = {103},
	url = {https://link.aps.org/doi/10.1103/PhysRevC.103.034601},
	doi = {10.1103/PhysRevC.103.034601},
	shorttitle = {Investigating high-energy proton-induced reactions on spherical nuclei},
	pages = {034601},
	number = {3},
	journaltitle = {Physical Review C},
	shortjournal = {Phys. Rev. C},
	publisher = {American Physical Society},
	author = {Fox, Morgan B. and Voyles, Andrew S. and Morrell, Jonathan T. and Bernstein, Lee A. and Lewis, Amanda M. and Koning, Arjan J. and Batchelder, Jon C. and Birnbaum, Eva R. and Cutler, Cathy S. and Medvedev, Dmitri G. and Nortier, Francois M. and O'Brien, Ellen M. and Vermeulen, Christiaan},
	urldate = {2024-12-31},
	date = {2021-03-01},
}

@misc{kingma_adam_2017,
	title = {Adam: A Method for Stochastic Optimization},
	url = {http://arxiv.org/abs/1412.6980},
	doi = {10.48550/arXiv.1412.6980},
	shorttitle = {Adam},
	number = {{arXiv}:1412.6980},
	publisher = {{arXiv}},
	author = {Kingma, Diederik P. and Ba, Jimmy},
	urldate = {2025-01-24},
	date = {2017-01-30},
	langid = {english},
	eprinttype = {arxiv},
	eprint = {1412.6980 [cs]},
}

@article{brzezinski_evaluation_1997,
	title = {Evaluation of 32Si as a tracer for measuring silica production rates in marine waters},
	volume = {42},
	rights = {© 1997, by the Association for the Sciences of Limnology and Oceanography, Inc.},
	issn = {1939-5590},
	doi = {10.4319/lo.1997.42.5.0856},
	pages = {856--865},
	number = {5},
	journaltitle = {Limnology and Oceanography},
	author = {Brzezinski, Mark A. and Phillips, Dennis R.},
	urldate = {2025-02-06},
	date = {1997},
	langid = {english},
}

@article{shusterman_surprisingly_2019,
	title = {The surprisingly large neutron capture cross-section of 88Zr},
	volume = {565},
	rights = {2019 Springer Nature Limited},
	issn = {1476-4687},
	url = {https://www.nature.com/articles/s41586-018-0838-z},
	doi = {10.1038/s41586-018-0838-z},
	pages = {328--330},
	number = {7739},
	journaltitle = {Nature},
	publisher = {Nature Publishing Group},
	author = {Shusterman, Jennifer A. and Scielzo, Nicholas D. and Thomas, Keenan J. and Norman, Eric B. and Lapi, Suzanne E. and Loveless, C. Shaun and Peters, Nickie J. and Robertson, J. David and Shaughnessy, Dawn A. and Tonchev, Anton P.},
	urldate = {2025-02-26},
	date = {2019-01},
	langid = {english},
}

@article{koning_tendl_2019,
	title = {{TENDL}: Complete Nuclear Data Library for Innovative Nuclear Science and Technology},
	volume = {155},
	issn = {0090-3752},
	url = {https://www.sciencedirect.com/science/article/pii/S009037521930002X},
	doi = {10.1016/j.nds.2019.01.002},
	series = {Special Issue on Nuclear Reaction Data},
	shorttitle = {{TENDL}},
	pages = {1--55},
	journaltitle = {Nuclear Data Sheets},
	shortjournal = {Nuclear Data Sheets},
	author = {Koning, A. J. and Rochman, D. and Sublet, J. -Ch. and Dzysiuk, N. and Fleming, M. and van der Marck, S.},
	urldate = {2025-02-27},
	date = {2019-01-01},
}

@article{chahrour_comparing_2022,
	title = {Comparing machine learning and interpolation methods for loop-level calculations},
	volume = {12},
	issn = {2542-4653},
	url = {https://scipost.org/10.21468/SciPostPhys.12.6.187},
	doi = {10.21468/SciPostPhys.12.6.187},
	pages = {187},
	number = {6},
	journaltitle = {{SciPost} Physics},
	shortjournal = {{SciPost} Phys.},
	author = {Chahrour, Ibrahim and Wells, James},
	urldate = {2025-03-14},
	date = {2022-06-08},
	langid = {english},
}

@article{trinchero_machine_2021,
	title = {Machine Learning Regression Techniques for the Modeling of Complex Systems: An Overview},
	volume = {10},
	issn = {2162-2272},
	url = {https://ieeexplore.ieee.org/abstract/document/9705310},
	doi = {10.1109/MEMC.2021.9705310},
	shorttitle = {Machine Learning Regression Techniques for the Modeling of Complex Systems},
	pages = {71--79},
	number = {4},
	journaltitle = {{IEEE} Electromagnetic Compatibility Magazine},
	author = {Trinchero, Riccardo and Canavero, Flavio},
	urldate = {2025-03-21},
	date = {2021},
}

@article{alizadeh_managing_2020,
	title = {Managing computational complexity using surrogate models: a critical review},
	volume = {31},
	issn = {1435-6066},
	url = {https://doi.org/10.1007/s00163-020-00336-7},
	doi = {10.1007/s00163-020-00336-7},
	shorttitle = {Managing computational complexity using surrogate models},
	pages = {275--298},
	number = {3},
	journaltitle = {Research in Engineering Design},
	shortjournal = {Res Eng Design},
	author = {Alizadeh, Reza and Allen, Janet K. and Mistree, Farrokh},
	urldate = {2025-03-24},
	date = {2020-07-01},
	langid = {english},
}

@book{van_rossum_python_2009,
	location = {Scotts Valley, {CA}},
	title = {Python 3 Reference Manual},
	isbn = {978-1-4414-1269-0},
	publisher = {{CreateSpace}},
	author = {Van Rossum, Guido and Drake, Fred L.},
	date = {2009-02},
}

@article{engle_recommended_2019,
	title = {Recommended Nuclear Data for the Production of Selected Therapeutic Radionuclides},
	volume = {155},
	issn = {0090-3752},
	url = {https://www.sciencedirect.com/science/article/pii/S0090375219300031},
	doi = {10.1016/j.nds.2019.01.003},
	series = {Special Issue on Nuclear Reaction Data},
	pages = {56--74},
	journaltitle = {Nuclear Data Sheets},
	shortjournal = {Nuclear Data Sheets},
	author = {Engle, J. W. and Ignatyuk, A. V. and Capote, R. and Carlson, B. V. and Hermanne, A. and Kellett, M. A. and Kibédi, T. and Kim, G. and Kondev, F. G. and Hussain, M. and Lebeda, O. and Luca, A. and Nagai, Y. and Naik, H. and Nichols, A. L. and Nortier, F. M. and Suryanarayana, S. V. and Takács, S. and Tárkányi, F. T. and Verpelli, M.},
	urldate = {2025-07-30},
	date = {2019-01-01},
}

@article{simmonds_design_1967,
	title = {Design of a nickel-63 electron absorption detector and analytical significance of high-temperature operation},
	volume = {39},
	issn = {0003-2700, 1520-6882},
	url = {https://pubs.acs.org/doi/abs/10.1021/ac60256a020},
	doi = {10.1021/ac60256a020},
	pages = {1428--1433},
	number = {12},
	journaltitle = {Analytical Chemistry},
	shortjournal = {Anal. Chem.},
	author = {Simmonds, Peter G. and Fenimore, David C. and Pettitt, Bruce C. and Lovelock, James E. and Zlatkis, Albert.},
	urldate = {2025-11-24},
	date = {1967-10-01},
	langid = {english},
}

@article{sgouros_radiopharmaceutical_2020,
	title = {Radiopharmaceutical therapy in cancer: clinical advances and challenges},
	volume = {19},
	rights = {2020 Springer Nature Limited},
	issn = {1474-1784},
	url = {https://www.nature.com/articles/s41573-020-0073-9},
	doi = {10.1038/s41573-020-0073-9},
	shorttitle = {Radiopharmaceutical therapy in cancer},
	pages = {589--608},
	number = {9},
	journaltitle = {Nature Reviews Drug Discovery},
	shortjournal = {Nat Rev Drug Discov},
	publisher = {Nature Publishing Group},
	author = {Sgouros, George and Bodei, Lisa and {McDevitt}, Michael R. and Nedrow, Jessie R.},
	urldate = {2025-11-24},
	date = {2020-09},
	langid = {english},
}

@article{tarkanyi_activation_2017,
	title = {Activation cross section data of proton induced nuclear reactions on lanthanum in the 34–65 {MeV} energy range and application for production of medical radionuclides},
	volume = {312},
	issn = {1588-2780},
	url = {https://doi.org/10.1007/s10967-017-5253-7},
	doi = {10.1007/s10967-017-5253-7},
	pages = {691--704},
	number = {3},
	journaltitle = {Journal of Radioanalytical and Nuclear Chemistry},
	shortjournal = {J Radioanal Nucl Chem},
	author = {Tárkányi, F. and Hermanne, A. and Ditrói, F. and Takács, S.},
	urldate = {2025-12-11},
	date = {2017-06-01},
	langid = {english},
}

@article{becker_cross_2020,
	title = {Cross section measurements for proton induced reactions on natural La},
	volume = {468},
	issn = {0168-583X},
	url = {https://www.sciencedirect.com/science/article/pii/S0168583X20300756},
	doi = {10.1016/j.nimb.2020.02.024},
	pages = {81--88},
	journaltitle = {Nuclear Instruments and Methods in Physics Research Section B: Beam Interactions with Materials and Atoms},
	shortjournal = {Nuclear Instruments and Methods in Physics Research Section B: Beam Interactions with Materials and Atoms},
	author = {Becker, K. V. and Vermeulen, E. and Kutyreff, C. J. and O’Brien, E. M. and Morrell, J. T. and Birnbaum, E. R. and Bernstein, L. A. and Nortier, F. M. and Engle, J. W.},
	urldate = {2025-12-11},
	date = {2020-04-01},
}

@article{morrell_measurement_2020,
	title = {Measurement of 139La(p,x) cross sections from 35–60 {MeV} by stacked-target activation},
	volume = {56},
	issn = {1434-601X},
	url = {https://doi.org/10.1140/epja/s10050-019-00010-0},
	doi = {10.1140/epja/s10050-019-00010-0},
	pages = {13},
	number = {1},
	journaltitle = {The European Physical Journal A},
	shortjournal = {Eur. Phys. J. A},
	author = {Morrell, Jonathan T. and Voyles, Andrew S. and Basunia, M. S. and Batchelder, Jon C. and Matthews, Eric F. and Bernstein, Lee A.},
	urldate = {2025-12-11},
	date = {2020-01-21},
	langid = {english},
}

@article{xiang_efficiency_2000,
	title = {Efficiency of generalized simulated annealing},
	volume = {62},
	url = {https://link.aps.org/doi/10.1103/PhysRevE.62.4473},
	doi = {10.1103/PhysRevE.62.4473},
	pages = {4473--4476},
	number = {3},
	journaltitle = {Physical Review E},
	shortjournal = {Phys. Rev. E},
	publisher = {American Physical Society},
	author = {Xiang, Y. and Gong, X. G.},
	urldate = {2025-12-22},
	date = {2000-09-01},
}

@article{xiang_generalized_1997,
	title = {Generalized simulated annealing algorithm and its application to the Thomson model},
	volume = {233},
	issn = {0375-9601},
	url = {https://www.sciencedirect.com/science/article/pii/S037596019700474X},
	doi = {10.1016/S0375-9601(97)00474-X},
	pages = {216--220},
	number = {3},
	journaltitle = {Physics Letters A},
	shortjournal = {Physics Letters A},
	author = {Xiang, Y and Sun, D. Y and Fan, W and Gong, X. G},
	urldate = {2025-12-22},
	date = {1997-08-25},
}

@article{xiang_generalized_2013,
	title = {Generalized Simulated Annealing for Global Optimization: The {GenSA} Package},
	volume = {5},
	issn = {2073-4859},
	url = {https://journal.r-project.org/archive/2013/RJ-2013-002/index.html},
	doi = {10.32614/RJ-2013-002},
	shorttitle = {Generalized Simulated Annealing for Global Optimization},
	pages = {13},
	number = {1},
	journaltitle = {The R Journal},
	shortjournal = {The R Journal},
	author = {Xiang, Yang and Gubian, Sylvain and Suomela, Brian and Hoeng, Julia},
	urldate = {2025-12-22},
	date = {2013},
	langid = {english},
}

@article{sobol_distribution_1967,
	title = {On the distribution of points in a cube and the approximate evaluation of integrals},
	volume = {7},
	issn = {0041-5553},
	url = {https://www.sciencedirect.com/science/article/pii/0041555367901449},
	doi = {10.1016/0041-5553(67)90144-9},
	pages = {86--112},
	number = {4},
	journaltitle = {{USSR} Computational Mathematics and Mathematical Physics},
	shortjournal = {{USSR} Computational Mathematics and Mathematical Physics},
	author = {Sobol', I. M},
	urldate = {2025-12-29},
	date = {1967-01-01},
}

@article{owen_scrambling_1998,
	title = {Scrambling Sobol' and Niederreiter–Xing Points},
	volume = {14},
	issn = {0885-064X},
	url = {https://www.sciencedirect.com/science/article/pii/S0885064X98904873},
	doi = {10.1006/jcom.1998.0487},
	pages = {466--489},
	number = {4},
	journaltitle = {Journal of Complexity},
	shortjournal = {Journal of Complexity},
	author = {Owen, Art B.},
	urldate = {2025-12-29},
	date = {1998-12-01},
}

@article{joe_constructing_2008,
	title = {Constructing Sobol Sequences with Better Two-Dimensional Projections},
	rights = {Copyright © 2008 Society for Industrial and Applied Mathematics},
	url = {https://epubs.siam.org/doi/10.1137/070709359},
	doi = {10.1137/070709359},
	journaltitle = {{SIAM} Journal on Scientific Computing},
	publisher = {Society for Industrial and Applied Mathematics},
	author = {Joe, Stephen and Kuo, Frances Y.},
	urldate = {2025-12-29},
	date = {2008-08-01},
	langid = {english},
}

@article{loh_latin_1996,
	title = {On Latin hypercube sampling},
	volume = {24},
	issn = {0090-5364, 2168-8966},
	url = {https://projecteuclid.org/journals/annals-of-statistics/volume-24/issue-5/On-Latin-hypercube-sampling/10.1214/aos/1069362310.full},
	doi = {10.1214/aos/1069362310},
	pages = {2058--2080},
	number = {5},
	journaltitle = {The Annals of Statistics},
	publisher = {Institute of Mathematical Statistics},
	author = {Loh, Wei-Liem},
	urldate = {2025-12-29},
	date = {1996-10},
}

@article{mckay_comparison_1979,
	title = {A Comparison of Three Methods for Selecting Values of Input Variables in the Analysis of Output from a Computer Code},
	volume = {21},
	issn = {0040-1706},
	url = {https://www.jstor.org/stable/1268522},
	doi = {10.2307/1268522},
	pages = {239--245},
	number = {2},
	journaltitle = {Technometrics},
	publisher = {[Taylor \& Francis, Ltd., American Statistical Association, American Society for Quality]},
	author = {{McKay}, M. D. and Beckman, R. J. and Conover, W. J.},
	urldate = {2025-12-29},
	date = {1979},
}

@article{stein_large_1987,
	title = {Large Sample Properties of Simulations Using Latin Hypercube Sampling},
	volume = {29},
	issn = {0040-1706},
	url = {https://www.tandfonline.com/doi/abs/10.1080/00401706.1987.10488205},
	doi = {10.1080/00401706.1987.10488205},
	pages = {143--151},
	number = {2},
	journaltitle = {Technometrics},
	publisher = {Taylor \& Francis},
	author = {Stein, Michael},
	urldate = {2025-12-29},
	date = {1987-05-01},
}

@article{hastie_surprises_2022,
	title = {Surprises in high-dimensional ridgeless least squares interpolation},
	volume = {50},
	issn = {0090-5364, 2168-8966},
	url = {https://projecteuclid.org/journals/annals-of-statistics/volume-50/issue-2/Surprises-in-high-dimensional-ridgeless-least-squares-interpolation/10.1214/21-AOS2133.full},
	doi = {10.1214/21-AOS2133},
	pages = {949--986},
	number = {2},
	journaltitle = {The Annals of Statistics},
	publisher = {Institute of Mathematical Statistics},
	author = {Hastie, Trevor and Montanari, Andrea and Rosset, Saharon and Tibshirani, Ryan J.},
	urldate = {2025-12-30},
	date = {2022-04},
}

@article{liang_just_2020,
	title = {Just Interpolate: Kernel “Ridgeless” Regression Can Generalize},
	volume = {48},
	issn = {0090-5364},
	url = {https://www.jstor.org/stable/26931513},
	shorttitle = {Just Interpolate},
	pages = {1329--1347},
	number = {3},
	journaltitle = {The Annals of Statistics},
	publisher = {Institute of Mathematical Statistics},
	author = {Liang, Tengyuan and Rakhlin, Alexander},
	urldate = {2025-12-30},
	date = {2020},
}

@article{ghorbani_linearized_2021,
	title = {Linearized two-layers neural networks in high dimension},
	volume = {49},
	issn = {0090-5364, 2168-8966},
	url = {https://projecteuclid.org/journals/annals-of-statistics/volume-49/issue-2/Linearized-two-layers-neural-networks-in-high-dimension/10.1214/20-AOS1990.full},
	doi = {10.1214/20-AOS1990},
	pages = {1029--1054},
	number = {2},
	journaltitle = {The Annals of Statistics},
	publisher = {Institute of Mathematical Statistics},
	author = {Ghorbani, Behrooz and Mei, Song and Misiakiewicz, Theodor and Montanari, Andrea},
	urldate = {2025-12-30},
	date = {2021-04},
}

@inproceedings{ansel_pytorch_2024,
	location = {New York, {NY}, {USA}},
	title = {{PyTorch} 2: Faster Machine Learning Through Dynamic Python Bytecode Transformation and Graph Compilation},
	volume = {2},
	isbn = {979-8-4007-0385-0},
	url = {https://dl.acm.org/doi/10.1145/3620665.3640366},
	doi = {10.1145/3620665.3640366},
	series = {{ASPLOS} '24},
	shorttitle = {{PyTorch} 2},
	pages = {929--947},
	booktitle = {Proceedings of the 29th {ACM} International Conference on Architectural Support for Programming Languages and Operating Systems, Volume 2},
	publisher = {Association for Computing Machinery},
	author = {Ansel, Jason and Yang, Edward and He, Horace and Gimelshein, Natalia and Jain, Animesh and Voznesensky, Michael and Bao, Bin and Bell, Peter and Berard, David and Burovski, Evgeni and Chauhan, Geeta and Chourdia, Anjali and Constable, Will and Desmaison, Alban and {DeVito}, Zachary and Ellison, Elias and Feng, Will and Gong, Jiong and Gschwind, Michael and Hirsh, Brian and Huang, Sherlock and Kalambarkar, Kshiteej and Kirsch, Laurent and Lazos, Michael and Lezcano, Mario and Liang, Yanbo and Liang, Jason and Lu, Yinghai and Luk, C. K. and Maher, Bert and Pan, Yunjie and Puhrsch, Christian and Reso, Matthias and Saroufim, Mark and Siraichi, Marcos Yukio and Suk, Helen and Zhang, Shunting and Suo, Michael and Tillet, Phil and Zhao, Xu and Wang, Eikan and Zhou, Keren and Zou, Richard and Wang, Xiaodong and Mathews, Ajit and Wen, William and Chanan, Gregory and Wu, Peng and Chintala, Soumith},
	urldate = {2026-01-05},
	date = {2024-04-27},
}

@article{bosso_application_2024,
	title = {Application of machine learning techniques to build digital twins for long train dynamics simulations},
	volume = {62},
	issn = {0042-3114},
	url = {https://doi.org/10.1080/00423114.2023.2174885},
	doi = {10.1080/00423114.2023.2174885},
	pages = {21--40},
	number = {1},
	journaltitle = {Vehicle System Dynamics},
	publisher = {Taylor \& Francis},
	author = {Bosso, N. and Magelli, M. and Trinchero, R. and Zampieri, N.},
	urldate = {2026-01-15},
	date = {2024-01-02},
}

@article{kudela_recent_2022,
	title = {Recent advances and applications of surrogate models for finite element method computations: a review},
	volume = {26},
	issn = {1433-7479},
	url = {https://doi.org/10.1007/s00500-022-07362-8},
	doi = {10.1007/s00500-022-07362-8},
	shorttitle = {Recent advances and applications of surrogate models for finite element method computations},
	pages = {13709--13733},
	number = {24},
	journaltitle = {Soft Computing},
	shortjournal = {Soft Comput},
	author = {Kudela, Jakub and Matousek, Radomil},
	urldate = {2026-01-15},
	date = {2022-12-01},
	langid = {english},
}

@article{cozad_learning_2014,
	title = {Learning surrogate models for simulation-based optimization},
	volume = {60},
	issn = {1547-5905},
	url = {https://onlinelibrary.wiley.com/doi/abs/10.1002/aic.14418},
	doi = {10.1002/aic.14418},
	pages = {2211--2227},
	number = {6},
	journaltitle = {{AIChE} Journal},
	publisher = {John Wiley \& Sons, Ltd},
	author = {Cozad, Alison and Sahinidis, Nikolaos V. and Miller, David C.},
	urldate = {2026-01-15},
	date = {2014},
	langid = {english},
}

@article{asher_review_2015,
	title = {A review of surrogate models and their application to groundwater modeling},
	volume = {51},
	rights = {© 2015. American Geophysical Union. All Rights Reserved.},
	issn = {1944-7973},
	url = {https://onlinelibrary.wiley.com/doi/abs/10.1002/2015WR016967},
	doi = {10.1002/2015WR016967},
	pages = {5957--5973},
	number = {8},
	journaltitle = {Water Resources Research},
	author = {Asher, M. J. and Croke, B. F. W. and Jakeman, A. J. and Peeters, L. J. M.},
	urldate = {2026-01-15},
	date = {2015},
	langid = {english},
}

@inproceedings{mccabe_multiple_2024,
	location = {Red Hook, {NY}, {USA}},
	title = {Multiple physics pretraining for spatiotemporal surrogate models},
	volume = {37},
	isbn = {979-8-3313-1438-5},
	series = {{NIPS} '24},
	pages = {119301--119335},
	booktitle = {Proceedings of the 38th International Conference on Neural Information Processing Systems},
	publisher = {Curran Associates Inc.},
	author = {{McCabe}, Michael and Blancard, Bruno Régaldo-Saint and Parker, Liam and Ohana, Ruben and Cranmer, Miles and Bietti, Alberto and Eickenberg, Michael and Golkar, Siavash and Krawezik, Geraud and Lanusse, Francois and Pettee, Mariel and Tesileanu, Tiberiu and Cho, Kyunghyun and Ho, Shirley},
	urldate = {2026-01-15},
	date = {2024-12-10},
}

@article{samadian_application_2025,
	title = {Application of Data-Driven Surrogate Models in Structural Engineering: A Literature Review},
	volume = {32},
	issn = {1886-1784},
	url = {https://doi.org/10.1007/s11831-024-10152-0},
	doi = {10.1007/s11831-024-10152-0},
	shorttitle = {Application of Data-Driven Surrogate Models in Structural Engineering},
	pages = {735--784},
	number = {2},
	journaltitle = {Archives of Computational Methods in Engineering},
	shortjournal = {Arch Computat Methods Eng},
	author = {Samadian, Delbaz and Muhit, Imrose B. and Dawood, Nashwan},
	urldate = {2026-01-15},
	date = {2025-03-01},
	langid = {english},
}

@article{khan_ai_2022,
	title = {{AI} and extreme scale computing to learn and infer the physics of higher order gravitational wave modes of quasi-circular, spinning, non-precessing black hole mergers},
	volume = {835},
	issn = {0370-2693},
	url = {https://www.sciencedirect.com/science/article/pii/S0370269322006396},
	doi = {10.1016/j.physletb.2022.137505},
	pages = {137505},
	journaltitle = {Physics Letters B},
	shortjournal = {Physics Letters B},
	author = {Khan, Asad and Huerta, E. A. and Kumar, Prayush},
	urldate = {2026-01-16},
	date = {2022-12-10},
}

@misc{loshchilov_decoupled_2019,
	title = {Decoupled Weight Decay Regularization},
	url = {http://arxiv.org/abs/1711.05101},
	doi = {10.48550/arXiv.1711.05101},
	number = {{arXiv}:1711.05101},
	publisher = {{arXiv}},
	author = {Loshchilov, Ilya and Hutter, Frank},
	urldate = {2026-01-17},
	date = {2019-01-04},
	langid = {english},
	eprinttype = {arxiv},
	eprint = {1711.05101 [cs]},
}

@article{khan_physics-inspired_2020,
	title = {Physics-inspired deep learning to characterize the signal manifold of quasi-circular, spinning, non-precessing binary black hole mergers},
	volume = {808},
	issn = {0370-2693},
	url = {https://www.sciencedirect.com/science/article/pii/S0370269320304317},
	doi = {10.1016/j.physletb.2020.135628},
	pages = {135628},
	journaltitle = {Physics Letters B},
	shortjournal = {Physics Letters B},
	author = {Khan, Asad and Huerta, E. A. and Das, Arnav},
	urldate = {2026-01-17},
	date = {2020-09-10},
}

@misc{misra_mish_2020,
	title = {Mish: A Self Regularized Non-Monotonic Activation Function},
	url = {http://arxiv.org/abs/1908.08681},
	doi = {10.48550/arXiv.1908.08681},
	shorttitle = {Mish},
	number = {{arXiv}:1908.08681},
	publisher = {{arXiv}},
	author = {Misra, Diganta},
	urldate = {2026-01-17},
	date = {2020-08-13},
	langid = {english},
	eprinttype = {arxiv},
	eprint = {1908.08681 [cs]},
}

@inproceedings{nair_rectified_2010,
	location = {Madison, {WI}, {USA}},
	title = {Rectified linear units improve restricted boltzmann machines},
	isbn = {978-1-60558-907-7},
	series = {{ICML}'10},
	pages = {807--814},
	booktitle = {Proceedings of the 27th International Conference on International Conference on Machine Learning},
	publisher = {Omnipress},
	author = {Nair, Vinod and Hinton, Geoffrey E.},
	urldate = {2026-01-17},
	date = {2010-06-21},
}

@article{fukushima_visual_1969,
	title = {Visual Feature Extraction by a Multilayered Network of Analog Threshold Elements},
	volume = {5},
	issn = {2168-2887},
	url = {https://ieeexplore.ieee.org/document/4082265},
	doi = {10.1109/TSSC.1969.300225},
	pages = {322--333},
	number = {4},
	journaltitle = {{IEEE} Transactions on Systems Science and Cybernetics},
	author = {Fukushima, Kunihiko},
	urldate = {2026-01-17},
	date = {1969-10},
}

@article{huber_robust_1964,
	title = {Robust Estimation of a Location Parameter},
	volume = {35},
	issn = {0003-4851, 2168-8990},
	url = {https://projecteuclid.org/journals/annals-of-mathematical-statistics/volume-35/issue-1/Robust-Estimation-of-a-Location-Parameter/10.1214/aoms/1177703732.full},
	doi = {10.1214/aoms/1177703732},
	pages = {73--101},
	number = {1},
	journaltitle = {The Annals of Mathematical Statistics},
	publisher = {Institute of Mathematical Statistics},
	author = {Huber, Peter J.},
	urldate = {2026-01-17},
	date = {1964-03},
}

@misc{meyer_alternative_2020,
	title = {An Alternative Probabilistic Interpretation of the Huber Loss},
	url = {http://arxiv.org/abs/1911.02088},
	doi = {10.48550/arXiv.1911.02088},
	number = {{arXiv}:1911.02088},
	publisher = {{arXiv}},
	author = {Meyer, Gregory P.},
	urldate = {2026-01-17},
	date = {2020-11-18},
	langid = {english},
	eprinttype = {arxiv},
	eprint = {1911.02088 [stat]},
}

@misc{apgar_117msn_2025,
	title = {117mSn and 119mTe Production via Proton Bombardment on Natural Antimony and Implications for Modeling Charged Particle Reactions},
	url = {http://arxiv.org/abs/2506.13948},
	doi = {10.48550/arXiv.2506.13948},
	number = {{arXiv}:2506.13948},
	publisher = {{arXiv}},
	author = {Apgar, Catherine E. and Voyles, Andrew S. and Batchelder, Jon C. and Cutler, Cathy S. and Fox, Morgan B. and Lee, Yun-Hsuan and Medvedev, Dmitri G. and Morrell, Jonathan T. and O'Brien, Ellen M. and Skulski, Michael and Vermeulen, C. Etienne and Bernstein, Lee A.},
	urldate = {2026-01-17},
	date = {2025-06-16},
	langid = {english},
	eprinttype = {arxiv},
	eprint = {2506.13948 [nucl-ex]},
}

@misc{klambauer_self-normalizing_2017,
	title = {Self-Normalizing Neural Networks},
	url = {http://arxiv.org/abs/1706.02515},
	doi = {10.48550/arXiv.1706.02515},
	number = {{arXiv}:1706.02515},
	publisher = {{arXiv}},
	author = {Klambauer, Günter and Unterthiner, Thomas and Mayr, Andreas and Hochreiter, Sepp},
	urldate = {2025-06-02},
	date = {2017-09-07},
	eprinttype = {arxiv},
	eprint = {1706.02515 [cs]},
}






\end{document}